\documentclass[submission, Proceedings]{SciPost}
\usepackage{calc}
\usepackage{graphicx}
\usepackage{float}
\usepackage{color}
\usepackage{amsmath}
\usepackage{subfig}
\usepackage{longtable}
\usepackage{graphics}
\usepackage{setspace}
\usepackage{appendix}

\def\be{\begin{equation}}
\def\ee{\end{equation}}
\def\ben{\begin{enumerate}}
\def\een{\end{enumerate}}
\def\bi{\begin{itemize}}
\def\ei{\end{itemize}}
\def\g2{{$(g-2)$}}

\begin{document}

\begin{center}{\Large \textbf{
The History of the Muon $(g-2)$ Experiments
}}\end{center}

\begin{center}
B. Lee Roberts ,
\end{center}

\begin{center}
   Department of Physics\\
  Boston University \\
  Boston, MA 02215  USA

* roberts@bu.edu
\end{center}

\begin{center}
\today
\end{center}

\definecolor{palegray}{gray}{0.95}
\begin{center}
\colorbox{palegray}{
  \begin{tabular}{rr}
  \begin{minipage}{0.05\textwidth}
    \includegraphics[width=8mm]{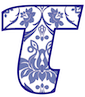}
  \end{minipage}
  &
  \begin{minipage}{0.82\textwidth}
    \begin{center}
    {\it Proceedings for the 15th International Workshop on Tau Lepton Physics,}\\
    {\it Amsterdam, The Netherlands, 24-28 September 2018} \\
    \href{https://scipost.org/SciPostPhysProc.1}{\small \sf scipost.org/SciPostPhysProc.Tau2018}\\
    \end{center}
  \end{minipage}
\end{tabular}
}
\end{center}


\section*{Abstract}
{\bf
I discuss the history of the muon $(g-2)$ measurements, beginning with the
Columbia-Nevis measurement that observed parity violation in muon decay,
and also
measured the muon $g$-factor for the first time, finding $g_\mu=2$.
The theoretical (Standard
Model) value contains contributions from quantum electrodynamics,
the strong interaction through
hadronic vacuum polarization and hadronic light-by-light loops, as well
as the electroweak contributions from the $W$, $Z$ and Higgs bosons.
The subsequent experiments,
first at Nevis and then with increasing precision at CERN,
measured the muon anomaly
$a_\mu = (g_\mu-2)/2$ down to a precision of 7.3 parts per million (ppm).
 The Brookhaven National Laboratory experiment
 E821 increased the precision to 0.54 ppm, and
observed for the first time the electroweak contributions.  Interestingly,
the value of $a_\mu$ measured at Brookhaven
appears to be larger than the Standard
Model value by greater than three standard deviations.  A new experiment,
Fermilab E989,  aims to improve on
 the precision by a factor of
four, to clarify whether this result is a harbinger of new physics entering
through loops, or from some experimental, statistical or systematic issue.
}

\vspace{10pt}
\noindent\rule{\textwidth}{1pt}
\tableofcontents\thispagestyle{fancy}
\noindent\rule{\textwidth}{1pt}
\vspace{10pt}

\section{Introduction}
\label{sec:intro}

The muon was first observed in cosmic rays by Paul Kunze~\cite{Kunze:1933} as
a ``particle of uncertain nature''\footnote{``Natur der oberen positiven
  Korpuskel nicht sicher bekannt.''}.  It was definitively identified by
Anderson and Neddemeyer~\cite{Anderson:1936zz}, and confirmed by Street and
Stevenson~\cite{Street:1937a,Street:1937me}, and by
Nishina et
al.,~\cite{Nishina:1937zz}.  There was significant confusion as to the nature
of this new particle.  It interacted too weakly with
matter~\cite{Conversi:1947ig} to be the
Yukawa particle~\cite{Yukawa:1938}, and it did not spontaneously decay to an
electron and a $\gamma$ ray~\cite{PhysRev.73.257,Lokanathan:1955},
nor did it convert to an
electron in the field of a nucleus~\cite{Lokanathan:1955}.  It became possible
that the muon might be like a heavy electron, which was a complete mystery.


\subsection{Spin and magnetic moments}

Our modern view of quantum mechanics of the leptons began with Dirac's famous
paper where he introduced the relativistic equation for the
electron~\cite{Dirac:1928hu}.  In that seminal paper
 he found that he was able to obtain the
measured magnetic moment of the electron:  ``an unexpected bonus for me,
completely unexpected''.~\cite{Pais:1998zz}

However, the story of spin began earlier. 
In an almost unknown paper, the idea of electron spin was first
proposed by A.H. Compton~\cite{Compton:1921sp},  who proposed
a spinning electron
to explain ferromagnetism, which he realized was difficult to explain
by any other means.  Subsequently  Unlenbeck and
Goudsmit~\cite{Uhlenbeck:1925nw,Uhlenbeck:1926na} proposed their spinning
electron to explain fine-structure splitting in atomic spectra.  However 
there was a factor of
two discrepancy between the measured fine structure splitting and that
predicted using Schr\"odinger quantum mechanics and their suggestion of spin.
This factor of two was
shown to be a relativistic effect 
 by L.H. Thomas~\cite{Thomas:1926dy,Thomas:1927yu}
which we now call  ``Thomas precession''.
Later, in a letter to Goudsmit, Thomas said~\cite{Thomas:1926sg}:
\begin{quote}
I think you and Uhlenbeck have been very lucky to get your spinning electron
published and talked about before Pauli heard of it.  It appears that more
than a year ago, Kronig believed in the spinning electron and worked out
something; the first person he showed it to was Pauli.  Pauli ridiculed the
whole thing so much that the first person became also the last and no one
else heard anything of it.  Which all goes to show that the infallibility of
the Deity does not extend to his self-styled vicar on earth.
\end{quote}
which adds a certain irony that we now talk about
the ``Pauli theory of spin''.
  
A spin 1/2 particle, has a magnetic moment along the spin:
\begin{equation}
\vec \mu = g \left(\frac{Qe}{2m}\right) \vec s\, ; \ \ {\rm where} \ \  g =
2(1+a)\,; \ \ {\rm or \ equivalently} \ \ 
a = \frac{g-2}{2}\, .
\end{equation}
I use the notation of Czarnecki and Mariano~\cite{Czarnecki:1900zz},
where $Q = \pm 1$ and $e>0$.  When placed in a magnetic field, there is a
torque on the spin, $\vec \mu \times \vec B$. If the particle is at rest,
the rate at which the spin
turns, the Larmor frequency, is given by
\begin{equation}
\vec \omega_S =\vec \omega_L \equiv g \left(\frac{Qe}{2m}\right)\vec B \,.
\end{equation}

\subsection{The first muon spin rotation experiments}

When Lee and Yang~\cite{Lee:1956qn} questioned whether the weak force
respected the parity symmetry, they laid out the details of several
experiments that could observe this violation including in muon decay, all of
which
were soon observed
experimentally~\cite{Wu:1957my,Garwin:1957hc,Friedman:1957mz}.  Parity
violation in pion decay produced polarized muons.  Furthermore, parity
violation in muon decays produced a correlation between the muon spin and the
highest energy positrons.   

Lee and Yang 
also pointed out that that parity violation would provide a way to measure
the muon magnetic moment. With the experimental observation of parity violation
in the  $\pi \rightarrow \mu \rightarrow e$ decay
  chain~\cite{Garwin:1957hc,Friedman:1957mz}, a tool to measure the magnetic
  moment became available.  The muon from pion decay at rest,
  $\pi^{\pm} \rightarrow  \mu^{\pm} + \bar \nu_\mu (\nu_\mu)$, is born
  100\% polarized to conserve angular momentum,
  since the neutrino (antineutrino) is left (right) handed. 
   For a beam of  pions, the very forward muons, as
  well as the very backward muons have a high degree of polarization.  
Thus the weak interaction provides experimentalists with information on where
the muon spin was initially
and in the decay, the highest energy positrons
are correlated with the muon spin.  Thus the polarized
muon at rest in a magnetic field
will precess, and the parity violating weak decay will cause a modulation of
the high-energy decay positrons with the Larmor frequency, $\omega_L$.

\begin{figure}[h!]
\begin{center}
\subfloat[]{
  \includegraphics[width=0.45\textwidth,angle=0]{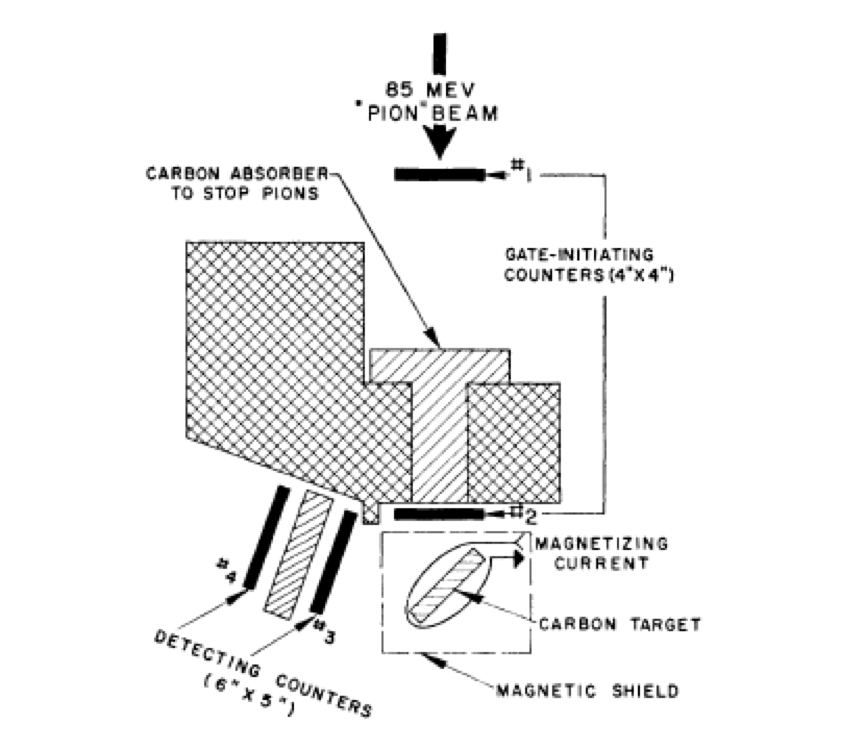}}
\subfloat[]{
 \includegraphics[width=0.45\textwidth,angle=0]{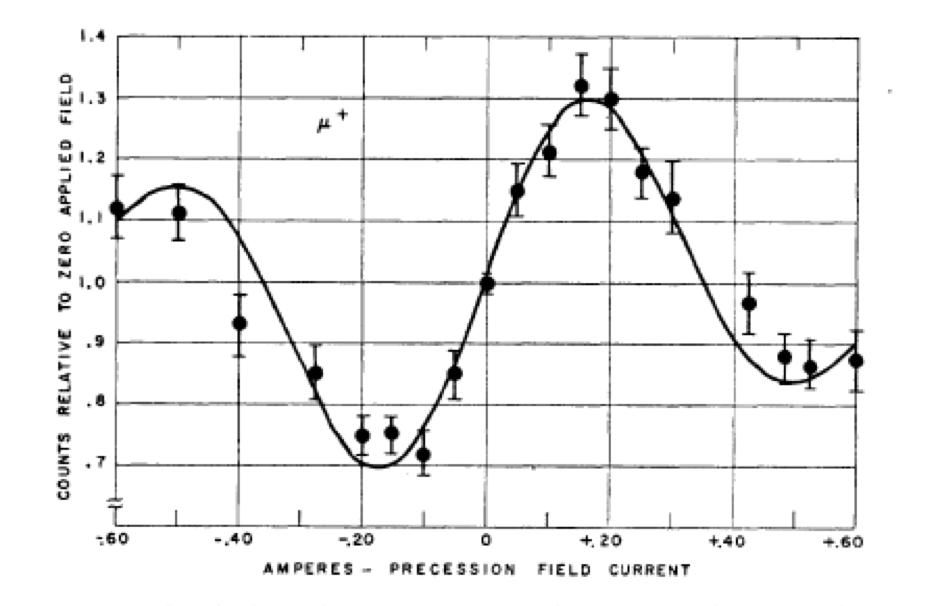}}
  \caption{(a) The experimental arrangement at Nevis. 85~MeV $\pi^+\mu^+$
    beam is incident. Four scintillation counters $1-4$
    defined an incident muon
    and a delayed muon decay.  The signal was  $1\cdot 2$, followed by a
    delayed $3\cdot 4$ coincidence within the time window of 0.75 - 2~$\mu$s.
    (b)The data from Garwin et al.,~\cite{Garwin:1957hc} showing the
    number of counts as a function of magnetic field during a narrow time
    window. (Reprinted with permission from Ref.~\cite{Garwin:1957hc}
    Copyright 1957 by the American Physical Society.) }
  \label{fg:Nevis}
\end{center}
\end{figure}

Garwin, Lederman and Weinrich~\cite{Garwin:1957hc} used a mixed beam of
$\pi^+$ and $\mu^+$ from the Nevis cyclotron.  Using a degrader, they stopped
the pions, and muons exiting the degrader were then stopped in a carbon
target, placed in a magnetic field as shown in Fig.~\ref{fg:Nevis}(a).
A scintillator telescope measured the
subsequent $\mu^+ \rightarrow e^+ \nu_\mu \bar \nu_e$ decay for a fixed time
interval. The magnetic field was varied, with higher field causing more spin
precession in the magnetic field before decay. The data from this first muon
spin rotation experiment~\cite{Garwin:1957hc}, {\it Observation of the
  Failure of  Conservation of Parity and Charge Conjugation in Meson
  Decays: the Magnetic moment of the Free Muon}, shown in
Fig.~\ref{fg:Nevis}(b)
permitted Garwin et al., to measure the muon
$g$-value, and found $g_\mu = 2 $ to 10\% precision.

\begin{figure}[h!]
\begin{center}
\subfloat[]{
  \includegraphics[width=0.455\textwidth,angle=0]{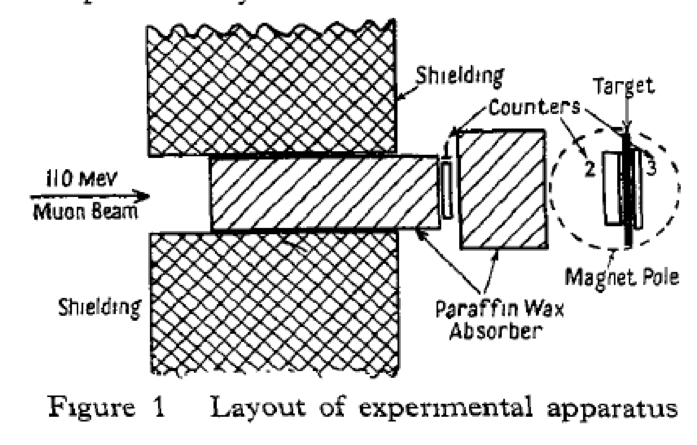}}
\subfloat[]{
 \includegraphics[width=0.54\textwidth,angle=0]{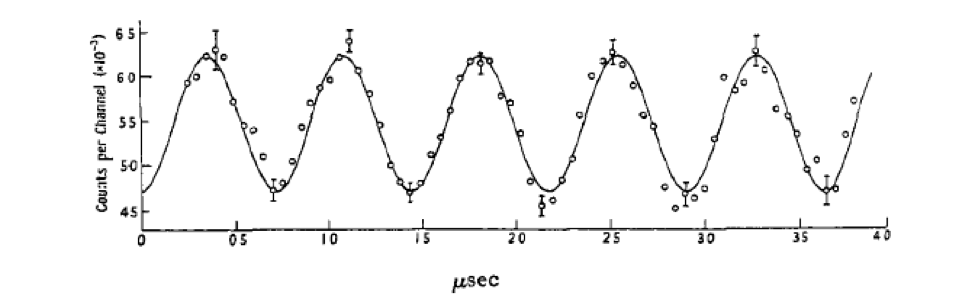}}
\caption{(a) The
    experimental set up at Liverpool. The incident muon beam is stopped in
    the target, $1 \cdot 2 \cdot \bar 3$ followed by a delayed $3\cdot \bar
    2$,
    (From Ref.~\cite{0370-1298-70-7-412}.
    Copyright IOP Publishing. 
     Reproduced with permission. All rights reserved.)
 (b) The muon time spectrum observed by Cassels et
    al.~\cite{0370-1298-70-7-412} as a function of time after the stopped
    muon, demonstrating the expected Larmor precession.
    The exponential muon lifetime was divided
    out using the measured muon lifetime.
    (From Ref.~\cite{0370-1298-70-7-412} Copyright IOP Publishing.
     Reproduced with permission. All rights reserved.)}
  \label{fg:Liverpool}
\end{center}
\end{figure}

Very soon thereafter, the precision on $g_\mu$ was
improved significantly by Cassels et al~\cite{0370-1298-70-7-412}
to  $g_\mu = 2.004 \pm 0.014$
at the University of Liverpool
cyclotron. Polarized muons were stopped in a Cu or C target,
which was inside of an external magnetic field, as shown in
Fig.~\ref{fg:Liverpool}(a).  A stopped muon was defined as $1\cdot 2 \cdot \bar 
3$. A delayed coincidence of $3\cdot \bar 2$ defined a muon decay. The stop
started an ``analyzer'', a voltage ramp, which was read out with a
multichannel analyzer when a muon decay trigger was received. This technique
was the beginning of what we now call 
a time to digital converter.  Their data shown in
Fig.~\ref{fg:Liverpool}(b) have the exponential muon lifetime  divided out,
using the measured muon lifetime from Bell and Hincks~\cite{PhysRev.84.1243}.

At the conclusion of this paper, the authors
state:
\begin{quote}the value of $g$ itself should be sought in a comparison of the
  precession  and cyclotron frequencies of muons in a magnetic field. The two
  frequencies are expected to differ only by the radiative correction. 
\end{quote}
We  explain this statement in the next section.

\subsection{Why go beyond $g=2$?}

In 1948 Julian Schwinger published a revolutionary paper on quantum
electrodynamics (QED) and the
magnetic moment of the electron~\cite{Schwinger:1948iu}.
In that paper he presented the very first  calculation of a
radiative correction in QED, which was motivated by the larger than 
expected hyperfine structure in hydrogen~\cite{Nafe:1947zz,Nagel:1947aa}.
Schwinger's result, the famous $\alpha/2 \pi$ mass-independent correction
 to the electron magnetic moment:
\begin{equation}
g_e = 2(1 + \frac{\alpha}{2\pi} ) = 2 (1 + 0.00116 \cdots ) \, ,
\end{equation}
 was the beginning of the calculation of
radiative corrections, which became increasingly important in the study
of lepton magnetic moments. This calculation was confirmed by the
famous experiment of Kush and Foley~\cite{Kusch:1948aa}. 
Today the electron anomaly has been measured to a few parts per
billion~\cite{Hanneke:2008tm}, which required the QED theory to be extended
to tenth-order (5 loops)~\cite{Aoyama:2017uqe}.

\begin{figure}[h!]
\begin{center}
  \includegraphics[width=0.9\textwidth,angle=0]{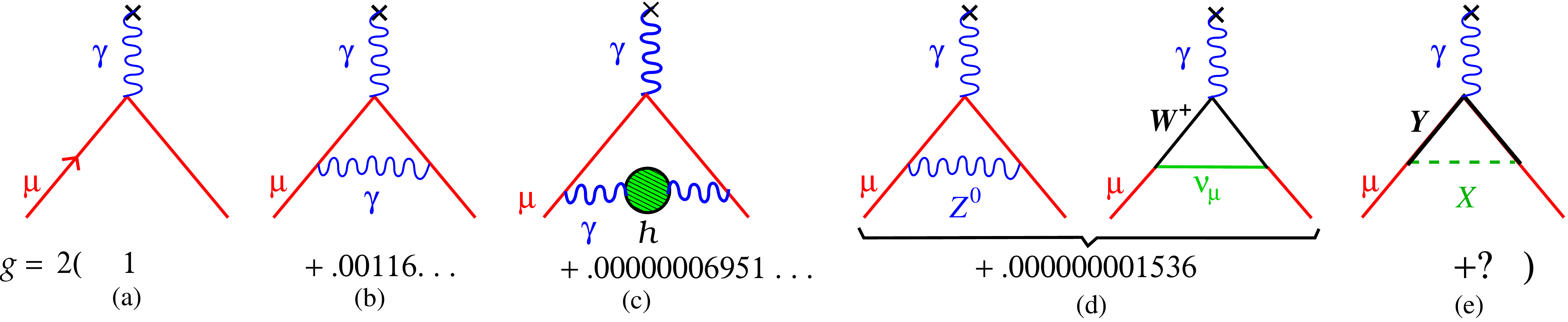}
  \caption{Radiative contributions to the muon anomaly from QED, the strong
    and the electroweak forces, showing their relative strengths.
    The Higgs boson does not contribute in the
    lowest order, but is important in the second-order electroweak
    contribution. (e)
    shows a possible contribution from as yet undiscovered new physics beyond
    the Standard Model. }
  \label{fg:feynman}
\end{center}
\end{figure}

For the muon, things are more
complicated, since the contribution of heavier physics, relative to the
electron, scales as $(m_\mu/m_e) \simeq 43,000.$ So the standard model value
of $a_\mu$ has measurable contributions from QED, the strong and electroweak
sector, as shown graphically in
Fig.~\ref{fg:feynman}.  Possible new, as yet undiscovered, physics beyond the
Standard Model could also contribute to the muon magnetic anomaly through
loops, as indicated in Fig.~\ref{fg:feynman}(e).

\subsection{Going beyond  $g_\mu =2$ in the laboratory}

In a series of follow-up experiments at Nevis using a stopped muon beam,
 Garwin, et
al.~\cite{Garwin:1960zz} improved on their initial measurement of the muon
magnetic moment. In a note in
proof, enabled by a new measurement of the muon mass, they obtained 
\begin{equation}
g_\mu = 2\left(1.00113^{+0.00016}_{-0.00012}\right) 
\end{equation}
which agreed with the value that
 Schwinger calculated~\cite{Schwinger:1948iu}. 
This final Nevis measurement  provided strong evidence that
in a magnetic field, the muon
behaved like a heavy electron.  This observation, combined with the failure
to observe the electromagnetic decay $\mu \rightarrow e + \gamma$
\cite{PhysRev.73.257,Lokanathan:1955}   clearly
pointed to the generation structure of the Standard Model.  The 
observation of the muon neutrino two years later~\cite{Danby:1962nd}
clarified the existence of the second generation, which became an important
ingredient in the Standard Model.

We should note at this point that these low-energy experiments with stopped
muons used positive muons, since low-energy negative muons
when brought into matter will stop and then be
 captured into atomic orbits.  Since the Bohr radius is proportional to
 the inverse of the mass of the orbiting particle, the muon quickly
gets inside of the atomic electron cloud and forms a hydrogen-like atom,
cascading down to the $1s$ atomic ground state of the muonic atom. It
will then undergo weak capture on the nucleus, or decay,
which will reduce the effective muon lifetime from 
the free muon lifetime of 2.2~$\mu$s.
So any measurement of the muon magnetic moment
that involves stopping a muon in matter, should be done with $\mu^+$. 

\subsection{The Spin Equations and Subsequent Experiments}

   When traversing a magnetic field where the velocity is
perpendicular to the magnetic field ($\vec \beta \cdot \vec B = 0$), 
the muon spin and momentum turn with the
frequencies
\begin{equation}
\vec  \omega_S = -g\frac{Qe}{2 m}\vec B - (1-\gamma) \frac{Qe}{\gamma m}\vec
B, \quad  
\vec  \omega_C = -\frac{Q e}{m \gamma}\vec B; \quad 
\vec \omega_a =\vec  \omega_S -\vec  \omega_C = - a_{\mu}\frac{Qe}{m}\vec B
\label{eq:diff-freq}
\end{equation}
where $\omega_a$ is the rate at which the spin turns relative to the
momentum. This difference frequency
provides a way to measure the anomaly, directly,
 and 
explains the quote from Cassels, et al.~\cite{0370-1298-70-7-412} above.
All subsequent $(g-2)$ experiments measured this difference frequency,
instead of measuring $g$ directly.  
Experimentally one measures $\omega_a$ and $\langle B\rangle$, where the
magnetic field is averaged over the muon distribution.

 Nuclear magnetic resonance (NMR) was used in all of the CERN experiments to
 measure 
 $\langle B\rangle$.  The presence of magnetic gradients over the NMR
 probes causes damping of the NMR signal, and reduces the precision of the NMR
 measurement. In addition,  if the magnetic field contains gradients, which are
necessary to focus the muon beam vertically, then the paths of the stored
muons needed to be known well, in order to calculate the average magnetic
field on the muon ensemble.   
As the precision on $a_\mu$ was increased in a
series of three experiments at CERN, it became necessary to come up with a
new way of focusing the beam, which eliminated the need for magnetic
gradients. The technique used in the third CERN experiment will be  discussed
below in some detail, since it formed the basis of the Brookhaven-based
experiment, E821, as well as the ongoing Fermilab experiment, E989.

\section{The CERN Experiments}

\subsection{CERN-1}

The first CERN experiment was carried out at the CERN synchrocylotron with a
beam of $\mu^+$~\cite{Charpak:1961mz,Charpak:1962zz}.
It should be noted that one of the motivations to press on
with a more precise measurement was to search for a breakdown of QED. A
cutoff of the muon propagator of energy $\Lambda m_\mu c^2$, would modify
the anomaly to be $a = (\alpha/2 \pi)[1 - (2/3)\Lambda^{-2}]$. So from the
beginning, the muon $(g-2)$ experiments were searching for evidence of
new physics.  

\begin{figure}[h!]
\begin{center}
  \includegraphics[width=\textwidth,angle=0]{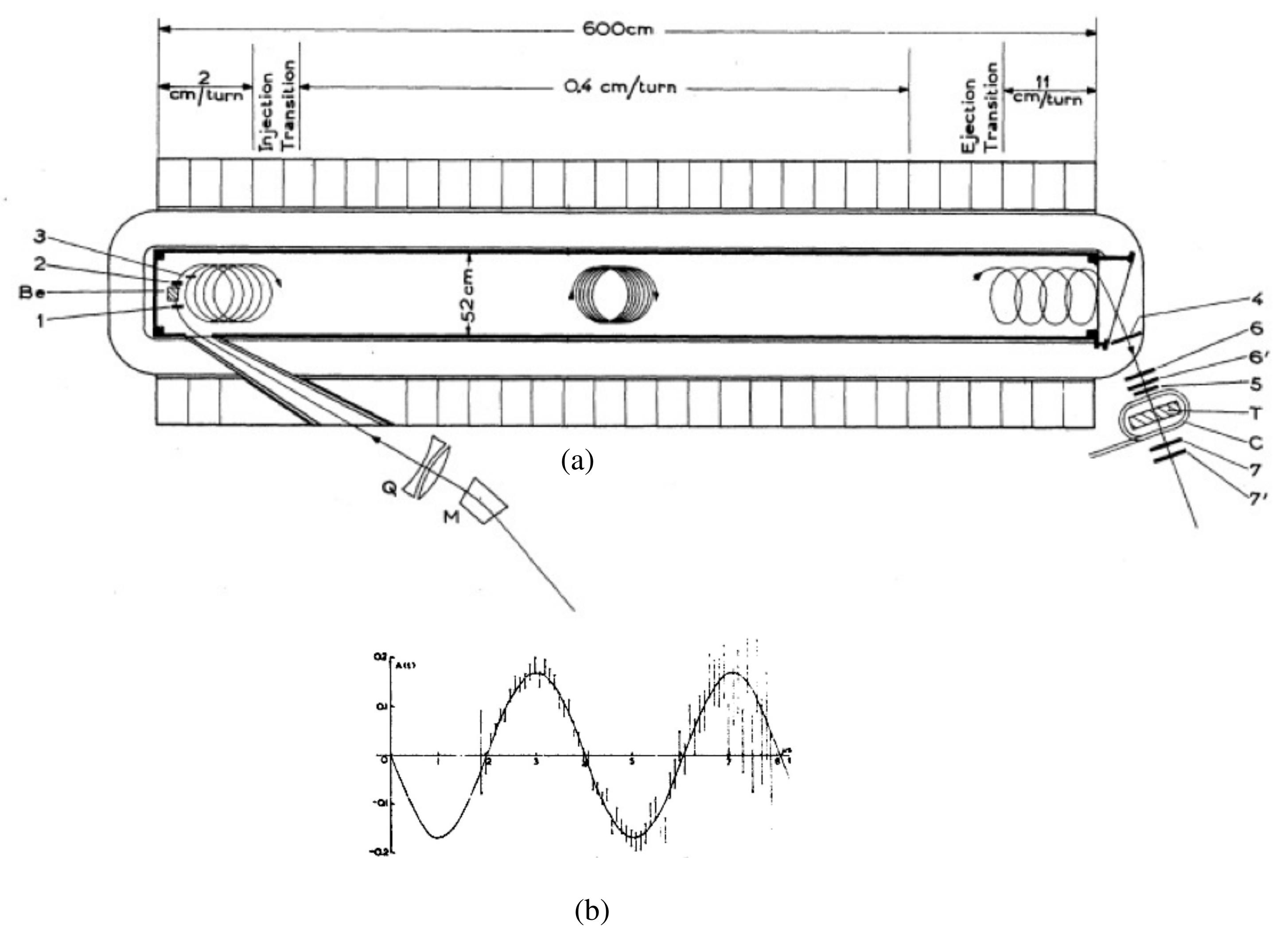} 
\caption{(a) The
  experimental set up for the first CERN $(g-2)$
  experiment~\cite{Charpak:1961mz}.  An incident $\mu^+$
  beam from the CERN SC was injected into a long dipole magnet, with the
  incident muon defined as a coincidence of the counters 1,2 and 3.  A beryllium
  degrader, placed between counters 2 and 3,
  was used to inject the beam into the magnet.  A large gradient at the end
  of the dipole was used to eject the beam and then it was stopped
  it in the methylene iodide target,
  as determined by  $1 \cdot 2 \cdot \bar 3$ followed by a
  delayed $4\cdot 5  \cdot 6 \cdot 6^\prime \cdot  \bar 7$
  followed by a $6\cdot 6^\prime$ or a $7\cdot7^\prime$ . 
  (Reprinted with permission from  Ref.~\cite{Charpak:1961mz}
  Copyright 1961 by the American Physical Society.)
  (b) The muon time spectrum from Charpak, et al.,~\cite{Charpak:1962zz}.
  The sinusoidal curve is the measured variation of the decay
  asymmetry with storage time, where the curve is the best fit. The time
  scale extends to 8~$\mu$s.  (From Ref.~\cite{Charpak:1962zz})
  \label{fg:CERN-1}}
\end{center}
\end{figure}

The beam was brought to a large dipole magnet with a
 magnetic field that contained a small gradient that caused the muon circular
 orbits to drift slowly toward the far end of the magnet as shown in
 Fig.~\ref{fg:CERN-1}(a).  The beam was injected into the magnet using a
 beryllium 
 degrader between the second and third scintillation counters.  At the end of
 the magnet, a large magnetic  gradient ejected the muons from the magnet.
 The exiting muons
 were stopped in a methylene iodide target, chosen because it did
 not destroy the muon polarization. The   backward (forward)  decays were
 recorded with a $6\cdot 6^\prime$ ( $7\cdot 7^\prime$) signal.
 The time distribution of the muon decays
 is shown in Fig.~\ref{fg:CERN-1}(b).    Note that the muons were
 non-relativistic, so the muon lifetime is essentially 2.2~$\mu$s.  
 The final result was\cite{Charpak:1962zz}:
 \begin{equation}
 a_\mu = (1162 \pm 5)\times 10^{-6}\, \
  \ (5164\ {\rm ppm)}\, .
 \end{equation}

The limitations of this experimental method are obvious: the muon lifetime at
rest is 2.2~$\mu$s, which limits the measurement period to a few
lifetimes. CERN-1 measured from 2~$\mu$s to 6.5~$\mu$s~\cite{Charpak:1961mz}
and then from 2~$\mu$s to 8~$\mu$s~\cite{Charpak:1962zz}.  
This short lifetime and measurement period means that
the number of muon decays measured is
small.  It became clear that to make a significant improvement in the
uncertainty, the experiment should be done
 with a more intense muon beam and at a higher muon
 energy, where the time dilated muon lifetime, $\gamma \tau_\mu$
 was significantly
 larger than 2.2~$\mu$s.

 \newpage

\subsection{CERN-2}

The next experiment at CERN was done using a storage ring.
The elements of a storage ring experiment are:
1) an incident beam of particles; 2) a
kick to store the injected
particles onto stable orbits in the storage ring;
3) detectors to detect the
daughter particles that come out of the storage ring reactions,
in this case the
positrons  from muon decay. The issue of injection into a storage
ring is quite complicated, and the latter two CERN experiments dodged this
technical problem by using the $\pi^+ \rightarrow \mu^+ + \nu_\mu$ decay as a
muon kicker to store muons.

\begin{figure}[h!]
\begin{center}
\subfloat[]{
  \includegraphics[width=0.5\textwidth,angle=0]{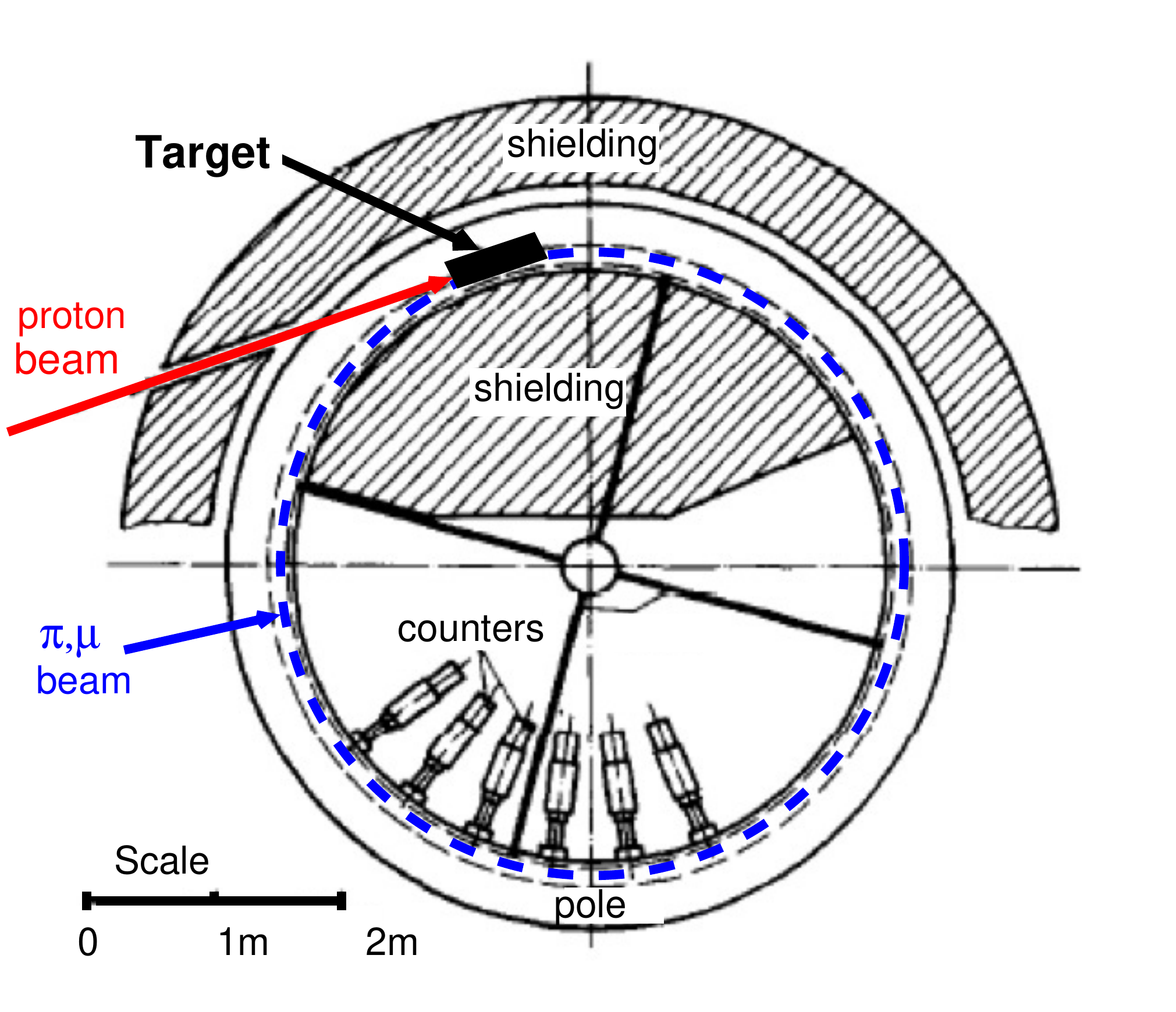}}
\subfloat[]{
 \includegraphics[width=0.5\textwidth,angle=0]{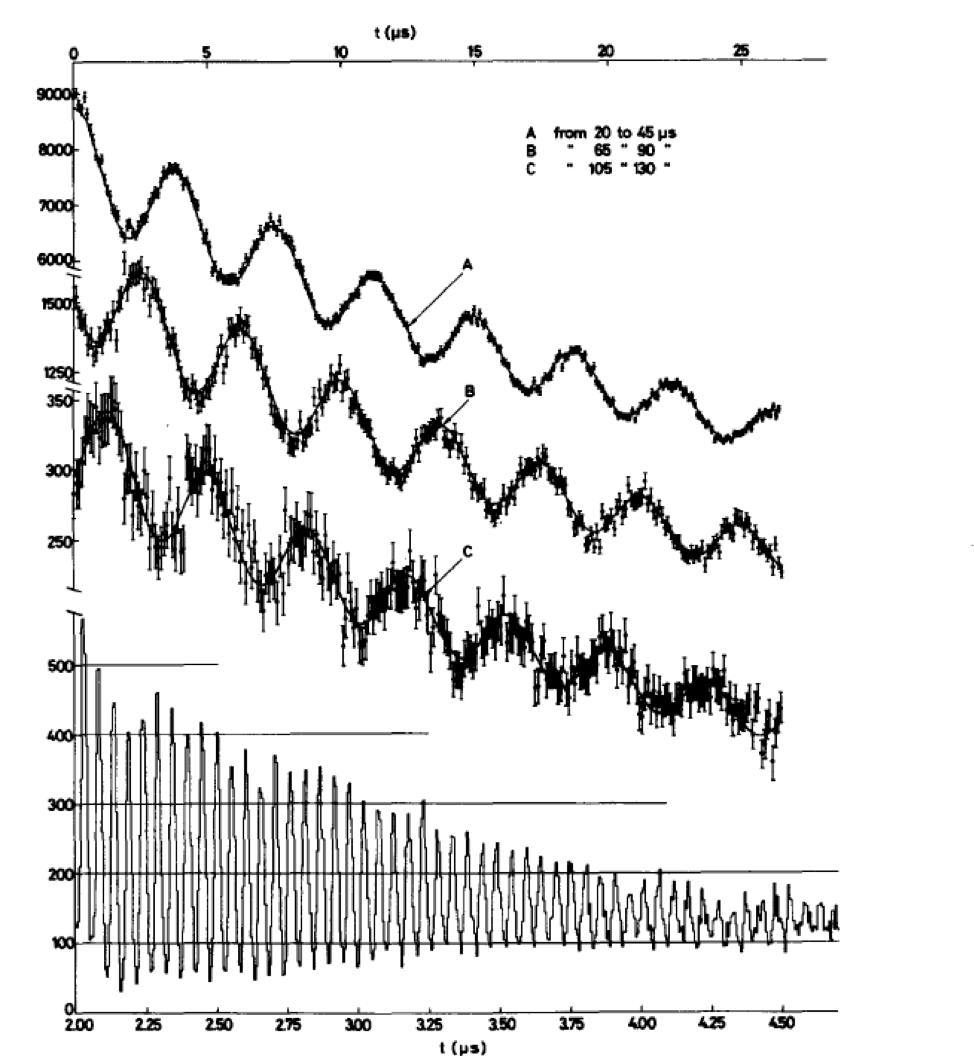}}
\caption{(a)The arrangement of the first CERN storage ring, showing the
  entering proton beam, the location of the shielding and detectors.
  There were four
  NMR probes (not shown) that were periodically inserted
  into the storage region to
  measure the magnetic field. (From Ref.~\cite{Bailey:1969pr} but
  re-labeled for  clarity.)  
  (b) The arrival time spectrum of high-energy positrons
  from the CERN-2 experiment. The rapid oscillations in the bottom of the
  figure are from the cyclotron frequency of the muon bunch, which dies out
  after 400 $\mu$s. The time spectrum is folded back on itself, where A is
  from 20-45 $\mu$s, B from 65-90 $\mu$s, C from 105-130 $\mu$s. (From
  Ref.~\cite{Bailey:1969pr}). 
  \label{fg:CERN-2}}
\end{center}
\end{figure}

The second CERN collaboration~\cite{Bailey:1969pr} built a 5~m diameter
magnetic weak focusing storage
ring that contained a pion production target inside of the
ring, as shown in Fig.~\ref{fg:CERN-2}(a).
The magnetic field was 1.711~T, $p_\mu = 1.27$~GeV/c and  $\gamma_\mu =12.06$.

The detectors were lead-scintillator
sandwiches optimized to measure the high-energy decay positrons,
which carry the muon spin information at the time of decay.  These detectors
 measured both the arrival time and (crudely) the energy of the decay
positrons. The arrival time spectrum is described by
\begin{equation}
  N(t) = N_0 e^{-{t}/{\gamma \tau_\mu}}\left[ 1 + {\mathcal A}\cos (\omega_a t + \phi)
    \right]\, , \ {\rm where} \
  \label{five-par}
\end{equation}
where $\omega_a$ was defined in Eq.~\ref{eq:diff-freq}, and ${\mathcal A}$
is the
product of the weak decay asymmetry $A$ times the average polarization
$\langle P \rangle$
of the muon beam.
The statistical error on $\omega_a$ is given by
\be
  \frac{\delta \omega_a}{\omega_a} = \frac{1}{\omega_a \gamma \tau_\mu}
  \sqrt{\frac{2}{NA^2\langle P\rangle^2}}\, .
  \label{eq:wa-error}
\ee

The result obtained in CERN-2 was
\begin{equation}
a_\mu^{\rm Exp} = (11661 \pm 3.1)\times 10^{-7}\, \ ( 266~{\rm ppm})
\end{equation}
which tested QED up to sixth-order, {\it viz.} to order $(\alpha/2\pi)^3$.

The large hadronic background from the pion production target placed
inside of the storage ring compromised the detectors, and limited the
sensitivity of this experiment. The statistical power was limited by the poor
efficiency for producing pions that were captured into an orbit in the
storage ring which then produce a stored muon.   

The limitation of this second CERN measurement was the small number of muons
stored, and the enormous background in the counters at injection caused by
the large hadronic flash in the detectors caused by  ``junk'' from the
production target.

\subsection{CERN-3}

To increase the precision of the measurement of $a_\mu$ further, one needed
to go to a higher muon lifetime, and to accumulate more data. The major issue
with a storage ring experiment that uses weak magnetic focusing, is with
magnetic gradients present. The major issue is
``How do you know the average field felt by the
muon distribution precisely without knowing the muon orbits precisely?''
The solution was found by the third CERN collaboration by
examining the the spin precession formula
for a muon in a storage ring with a uniform magnetic field  and an  electric
quadrupole 
field to provide vertical focusing.  In atomic physics this arrangement
is called a Penning Trap.

With the presence of both electric and magnetic fields,
the cyclotron frequency becomes
\begin{equation}
  \vec  \omega_C =- \frac{Qe}{m}
  \left[\frac{\vec B}{ \gamma} 
- \frac{\gamma}{ \gamma^2 -1} 
\left(\frac{ \vec \beta \times \vec E }{ c}\right)
\right],
\label{eq:cyc-E}
\end{equation}

and the spin rotation frequency is
\begin{equation}
\vec \omega_S = - \frac{Qe}{ m }
\left[
  \left(
         \frac{g }{ 2} -1 + \frac{1}{ \gamma} \right) \vec B
         - \left(\frac{g }{ 2} -1 \right)
         \frac{\gamma }{ \gamma + 1}(\vec \beta \cdot \vec B)\vec \beta
         -
  \left(
\frac{g }{ 2} - \frac{\gamma }{ \gamma + 1}
\right) 
\left(\frac{ \vec \beta \times \vec E }{ c}\right)
\right]\, ,
\label{eq:spin-freq}
\end{equation}
where the $\vec \beta \cdot \vec B$ term comes from the vertical pitching
motion of
the muons in the weak focusing storage ring.

Eq.~\ref{eq:spin-freq}
was first discovered by Thomas~\cite{Thomas:1927yu} in 1927. We use
the version given in Jackson's book~\cite{jackson-classical-1999}
in modern notation, which is equivalent to Thomas' Eq. 4.121 in
Ref.~\cite{Thomas:1927yu}.
\footnote{Bargmann, Michel and 
Telegdi~\cite{Bargmann:1959gz} also studied this
problem.}

Using  $a_{\mu} = (g_{\mu} -2)/2$, we find that
the spin difference frequency
is\footnote{Strictly speaking, the rate of change of the angle between the
  spin and the 
momentum vectors,
$|\vec\omega_{a_\mu}|$= `precession frequency',
is equal to $|\vec\omega_{diff}|$ only if
$\vec\omega_S$
and $\vec \omega_C$ are parallel. For the E821 and E989 experiments,
the angle between $\vec\omega_S$ and $\vec\omega_C$ is always small and the
rate of oscillation of $\vec\beta$ out of pure circular motion is fast compared
to $\omega_{a_\mu}$, allowing us in the following discussion
the make the approximation that $\vec\omega_{a_\mu}\simeq\vec\omega_{diff}$.
More general calculations, where this approximation is not made, are found in
References~\cite{Farley:1972zy,Combley:1974tw,Field:1974pe,Farley:1990gm}. 
For the CERN-3, E821 and E989 experimental conditions, the results
presented here are the same as those in these references. }
\begin{equation}
\vec \omega_{diff}=  \vec \omega_S -\vec\omega_C     \simeq \omega_{a_\mu} 
=  - \frac{Qe}{ m}
\left[ a_{\mu} \vec B -  
a_{\mu}\left( {\gamma \over \gamma + 1}\right)
(\vec \beta \cdot \vec B)\vec \beta 
- \left( a_{\mu}- {1 \over \gamma^2 - 1} \right) 
{ {\vec \beta \times \vec E }\over c }\right]\,.
\label{eq:Ediffreq}
\end{equation}

For the moment, we ignore the $(\vec \beta \cdot \vec B)$ term in
Eq.~\ref{eq:Ediffreq} and get
\begin{equation}
\omega_a =  - \frac{Qe}{ m} \left[ a_{\mu} \vec B -
  \left( a_{\mu}- {1 \over \gamma^2 - 1} \right) 
{ {\vec \beta \times \vec E }\over c }\right]\,.
\end{equation}
While a relativistic particle moving in an electric field will experience a
motional magnetic field, the negative sign in the parentheses introduces a
cancellation at $\gamma_m = 29.3$, $p_\mu = 3.09$~GeV/c.
By building a muon storage ring that
operates at the ``magic'' $\gamma$, the effect of the motional magnetic field
is minimized. Small corrections are necessary to account for the vertical
pitching motion, the ($\vec \beta \cdot \vec B$) term, and for the
fact that not all muons are at the central radius, and therefore not at magic
momentum. For details see~\cite{Grange:2015fou,Miller:2018jum,Field:1974pe}
and references therein.

\begin{figure}[h!]
\begin{center}
\subfloat[]{
  \includegraphics[width=0.5\textwidth,angle=0]{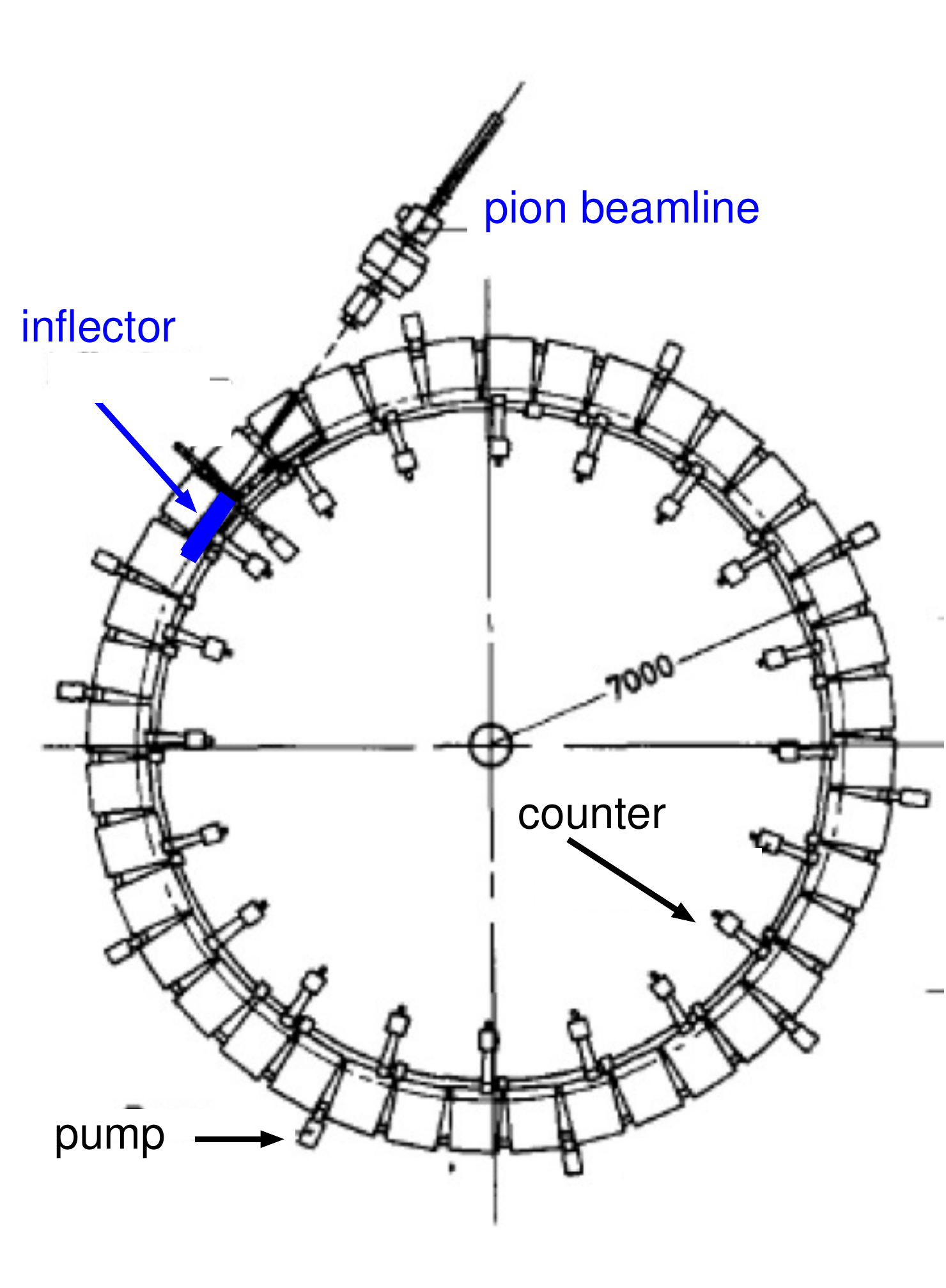}}
\subfloat[]{
 \includegraphics[width=0.5\textwidth,angle=0]{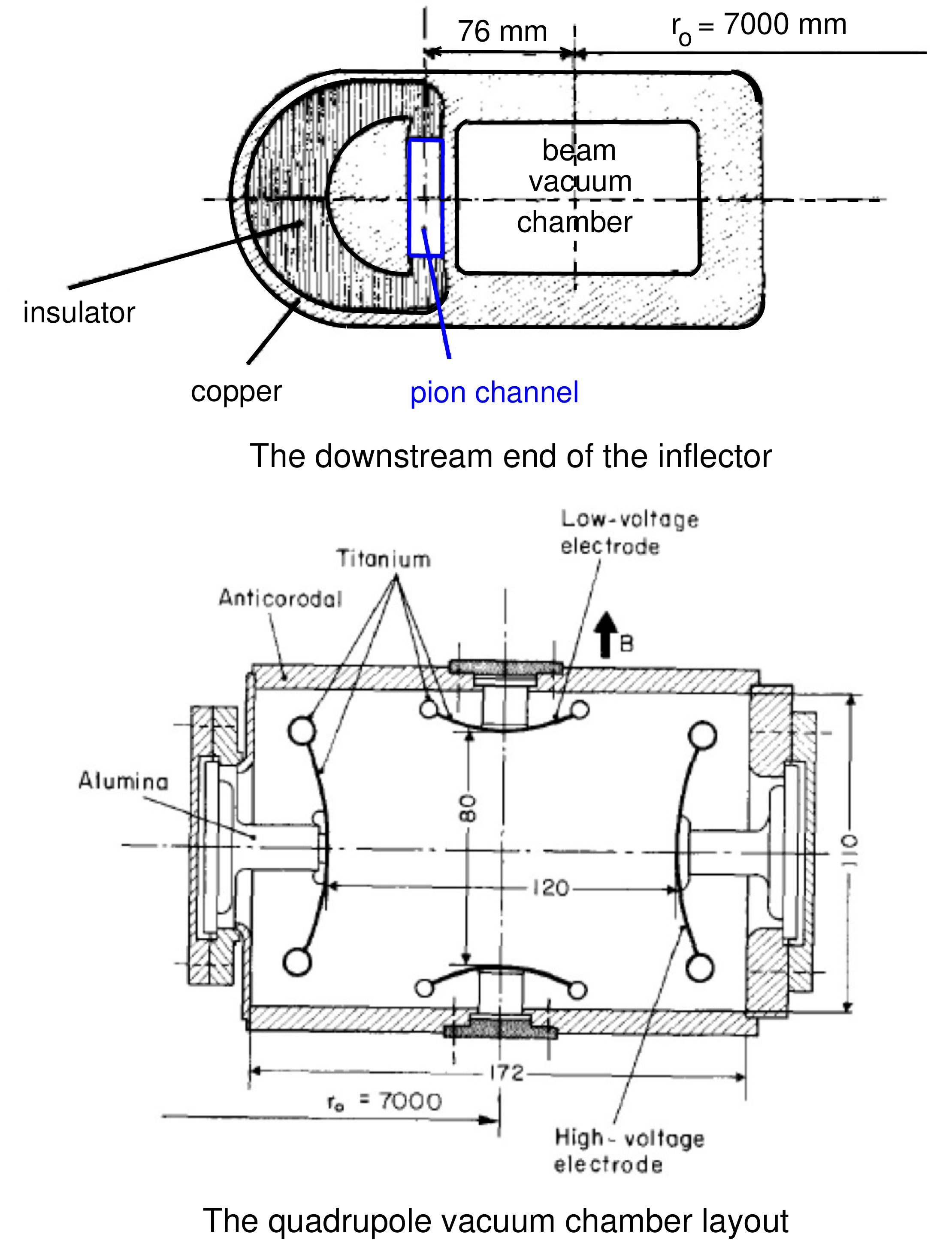}}
\caption{
  (a)A plan view of the 7,000~mm diameter
  CERN-3 storage ring, showing the 40-magnet ring.
  (From Ref.~\cite{Bailey:1978mn} re-labeled for
  clarity.) 
  (b) Upper figure: The downstream end of  the pulsed inflector magnet
  of  the CERN-3 experiment. (From Ref.~\cite{Bailey:1978mn} re-labeled for
  clarity.)  Lower figure: The  electric quadrupole
  arrangement inside of the rectangular vacuum chamber profile.
  (From Ref.~\cite{Bailey:1978mn}.) 
  \label{fg:CERN-3}}
\end{center}
\end{figure}

A new 7~m diameter storage ring,  composed of 40
contiguous magnets,
was designed and constructed as shown in
Fig.~\ref{fg:CERN-3}(a). Twenty four lead-scintillator shower counters were
placed symmetrically around the inside of the ring to measure the arrival
time and energy 
of the positrons.

\begin{figure}[h!]
\begin{center}
\subfloat[]{
  \includegraphics[width=0.5\textwidth,angle=0]{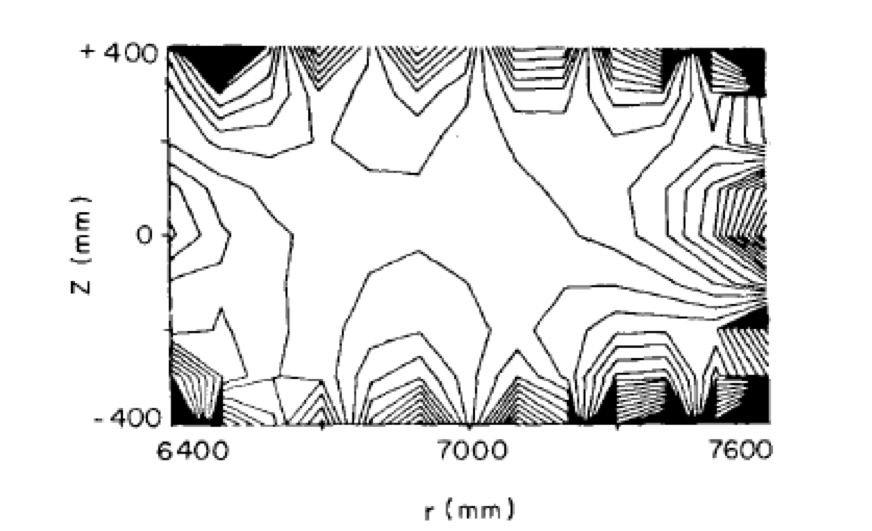}}
\subfloat[]{
 \includegraphics[width=0.5\textwidth,angle=0]{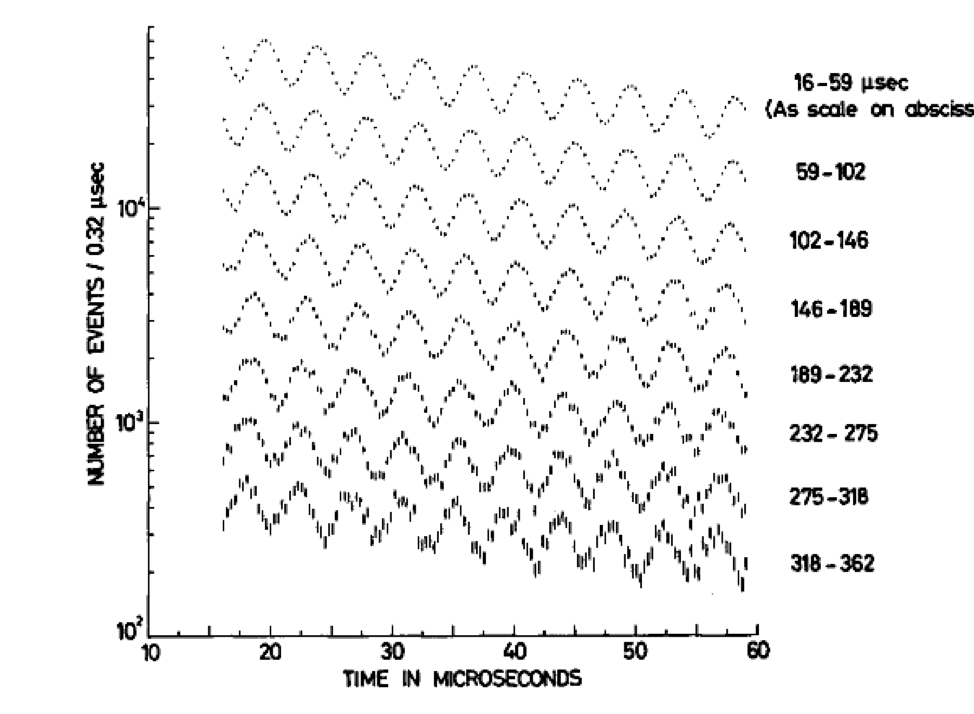}}
\caption{
  (a)  The magnetic field averaged over azimuth from the CERN-3
  experiment. The contours are 2 ppm (3~$\mu$T). The periodic structure of the
  contours most likely came from the width of the grinding wheel used in the
  shimming.   (From Ref.~\cite{Bailey:1978mn}).
  (b)The arrival time spectrum from a portion of the
  CERN-3 data. (From Ref.~\cite{Bailey:1975qi})
  \label{fg:CERN-3-field-data}}
\end{center}
\end{figure}

There were four significant improvements over the CERN-2
experiment: 1) The injection of a pion beam into the storage ring;  2) The
development of an inflector magnet that canceled the field of the
storage ring  so that the beam
deflection  entering  into
the storage ring was minimal; 3) The use of the magic $\gamma = 29.3$ meant
that the muon lifetime was extended to 64.4~$\mu$s, which significantly
increased the measurement time; 4). The more uniform magnetic field with weak
electric focusing made it easier to determine the average magnetic field
weighted over the muon distribution. The magnetic field averaged over azimuth
is shown in Fig.~\ref{fg:CERN-3-field-data}(a).

The inflector
magnetic field was generated by a current pulse which rose to a
peak value of 300~kA in 12~$\mu$s~\cite{Bailey:1978mn}.
 Electrostatic
quadrupoles~\cite{Flegel:1973zq} with 2-fold symmetry provided weak vertical
focusing. The quadrupoles were pulsed on during the data collection time, and
then off between fills of the ring to minimize the trapping of electrons by
the quadrupole field, to minimize sparks in the system.

The time spectrum of high-energy positrons is shown in
Fig.~\ref{fg:CERN-3-field-data}(b). The CERN-3 experiment
measured both positive and
negative muons, and the final results were~\cite{Bailey:1978mn}
\begin{eqnarray}
a_{\mu^+} &=& (1\,165\,911 \pm 11)\times 10^{-9} \ {\rm (10~ ppm)}\\
  a_{\mu^-} &=& (1\,165\,937 \pm 12)\times 10^{-9}  \ {\rm (10~ ppm)}\\
    a_\mu \  &=& (1\,165\,924 \pm 8.5)\times 10^{-9}  \ {\rm (7.3~ ppm)}
\end{eqnarray}
which agreed well with the Standard Model Value. To compare with theory, it
became necessary to include QED to sixth order, and the hadronic
contribution. The precision was not adequate to be sensitive to
eighth-order QED, or the electroweak contribution.

\section{Brookhaven Experiment E821}

To increase the precision on the muon anomaly, with the goal
to observe the electroweak
contribution, as well as to search for contributions from New Physics such as
supersymmetry, a new experiment was proposed for the Brookhaven National
Laboratory (BNL) Alternating Gradient Synchrotron (AGS).  The first
meetings began at Brookhaven around 1984. 
The plan was to use the magic $\gamma_m = 29.3$ and electrostatic
focusing, that was developed at CERN. A collaboration was formed,
that eventually became BNL Experiment E821. Final approval came from the
U.S. Department of Energy and from the Laboratory in 1989.

There were a number of significant improvements planned for the new
experiment: 1) A much more uniform magnetic field using a superferric
superconducting storage ring magnet; 2) A passive
superconducting inflector magnet; 3) A beam tube NMR trolley that could map
the magnetic field in the storage ring often, by simply turning the muon beam
off; 4) An array of 378 NMR probes around the ring to  continuously monitor
the magnetic field during data collection;
5) A 4-fold symmetry for the electrostatic quadrupoles instead of
2-fold, which resulted in $\sqrt{\beta_{max}/\beta_{min}} = 1.04$, compared to
3.26 for the two-fold symmetry used at CERN; 6) A circular beam with
collimators to minimize the importance of both
higher magnetic multipoles and higher moments of the muon 
beam distribution in the determination of the average magnetic field felt by
the muons;  7) A fast muon kicker to directly inject  muons
into the storage ring rather than pions,
which would drastically reduce the flash from the beam
pions after injection; and 8) A much more intense
pion/muon beam than was available at CERN.

\begin{figure}[h!]
\begin{center}
\subfloat[]{
  \includegraphics[width=0.37\textwidth,angle=0]{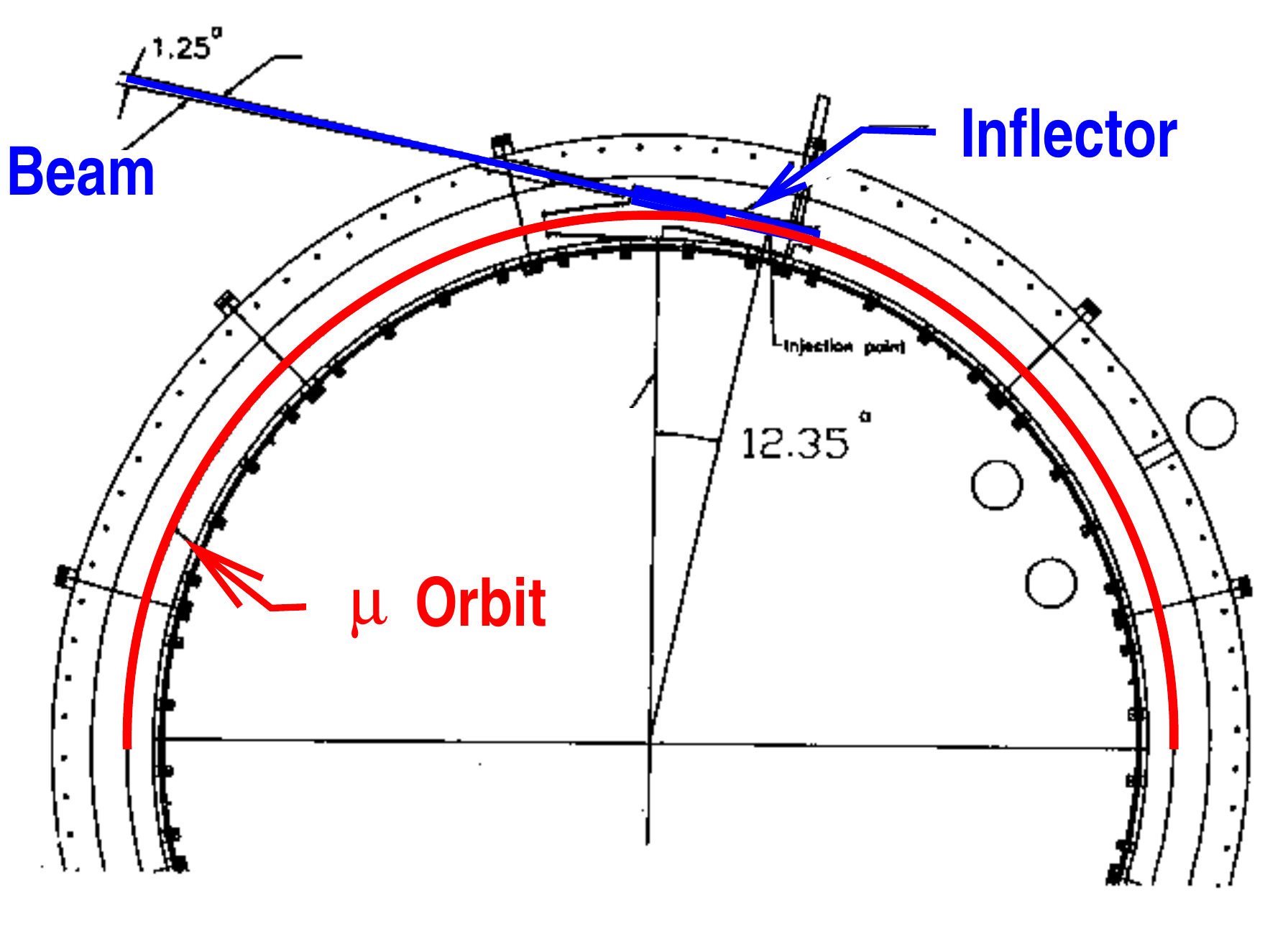}}
\subfloat[]{
 \includegraphics[width=0.63\textwidth,angle=0]{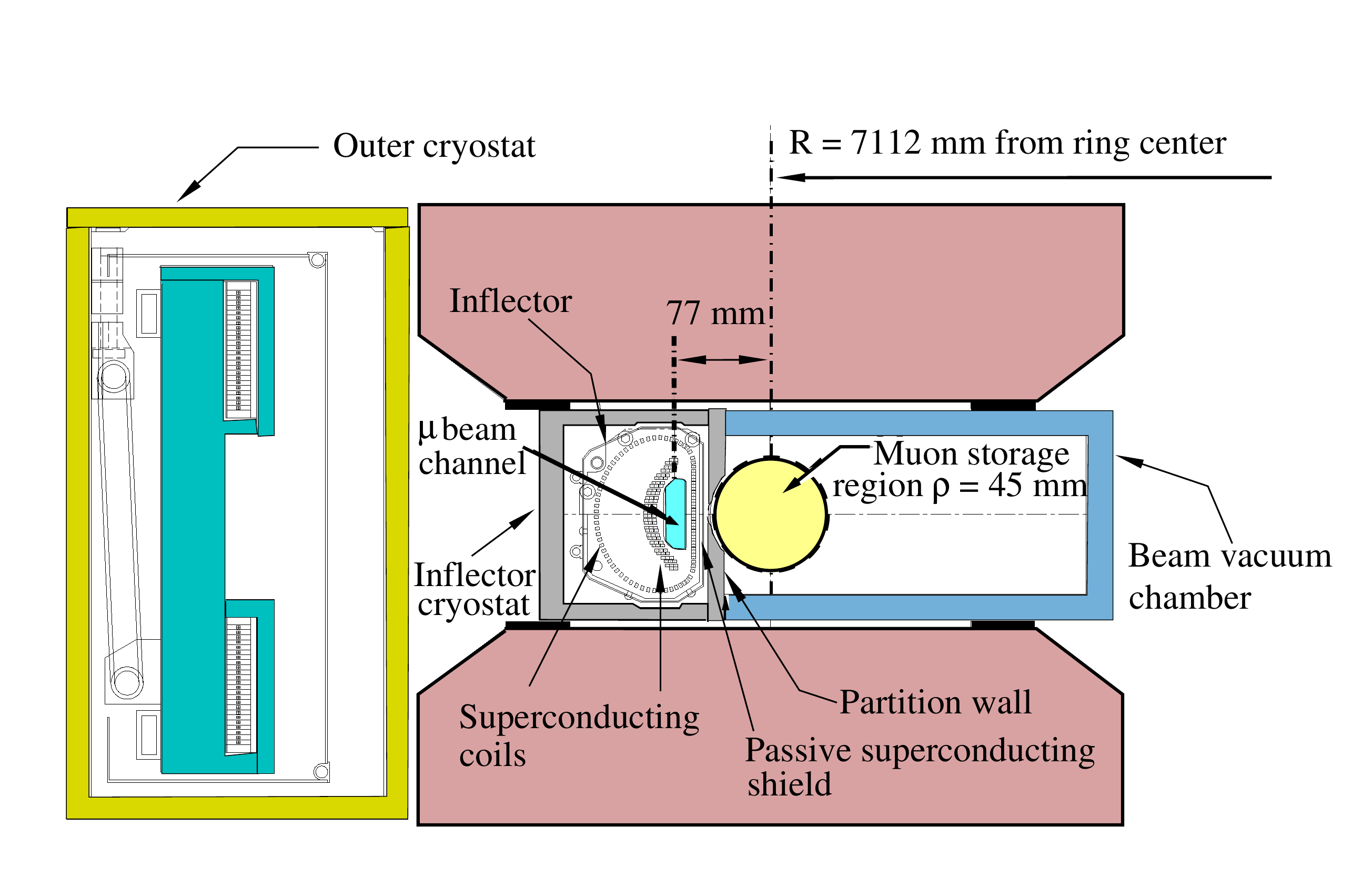}}
\caption{
  (a) A plan view of the incoming beam showing the location of the
  1.7~m long superconducting inflector.
  (b) An elevation view of the inflector exit.  The muon
  beam is into the page. The current in the ``C'' shaped arrangement of
  conductors flows in one direction, and in the opposite direction in the
  backward ``D'' shaped coil, producing a  uniform vertical
  dipole field opposite to the main magnet field of 1.45~T,
  so that the beam enters
  the storage ring undeflected.
   \label{fg:BNL-injection-geo}}
\end{center}
\end{figure}

The E821 beamline was designed to permit either pion injection or muon
injection into the storage ring.  A 24~GeV/c momentum proton pulse ($\sigma
\simeq 25$~ns)  from the
AGS containing up to $7 \times 10^{12}$ protons,
 was extracted from the AGS and brought to a pion production
target. The secondary beamline contained an 80~m pion decay channel, which had
momentum selection collimators both before and after the decay
channel~\cite{Bennett:2006fi}. After the final momentum selection, the beam
was brought into the experimental hall and entered the storage ring through a
penetration in the the yoke shown in Fig~\ref{fg:BNL-injection-geo}(a).

Before the fast muon kicker became available,
E821 had a short pion injection run in the style of CERN-3 using the $\pi
\rightarrow \mu$ decay to provide the kick. The initial
flash in the electron calorimeters from the injected pion beam was enormous.
This flash prevented us from analyzing data
before 75~$\mu$s after injection~\cite{Carey:1999dd}, more
than one muon 
lifetime of 64~$\mu$s.

When we switched to muon
injection, things became much more
manageable.   Once the fast muon kicker became
available, the muon beam  beam entering the storage ring had a pion
to muon ratio $\simeq 1:1$.  

We were able to begin the fit around 30~$\mu$s after injection,
which was after the muon beam debunched in the ring, and the initial flash
from the pions in the muon beam died away. Muon injection enabled us
 to improve on the precision of $a_\mu$ by a
factor of 14 over the CERN-3 result.  
  Fig~\ref{fg:BNL-injection-geo}(a) shows the geometry of the
incoming muon beam, and (b) the region at the  inflector exit.

A superconducting inflector magnet was designed
 based on a truncated double cosine
 theta design~\cite{Krienen:1989ba}.  This design provided 
 a superconducting septum magnet
that had minimal magnetic flux leakage into the muon storage
region. Furthermore it was surrounded with a passive superconducting shield,
that essentially eliminated the flux leakage from the inflector
magnet~\cite{Yamamoto:2002bb}.   This static superconducting
design eliminated the repetition rate limitations of the pulsed
CERN-3 inflector,
and had the important advantage that it did not produce any eddy currents or
other transient magnetic fields associated with the beam injection.

\begin{figure}[h!]
\begin{center}
\subfloat[]{
  \includegraphics[width=0.5\textwidth,angle=0]{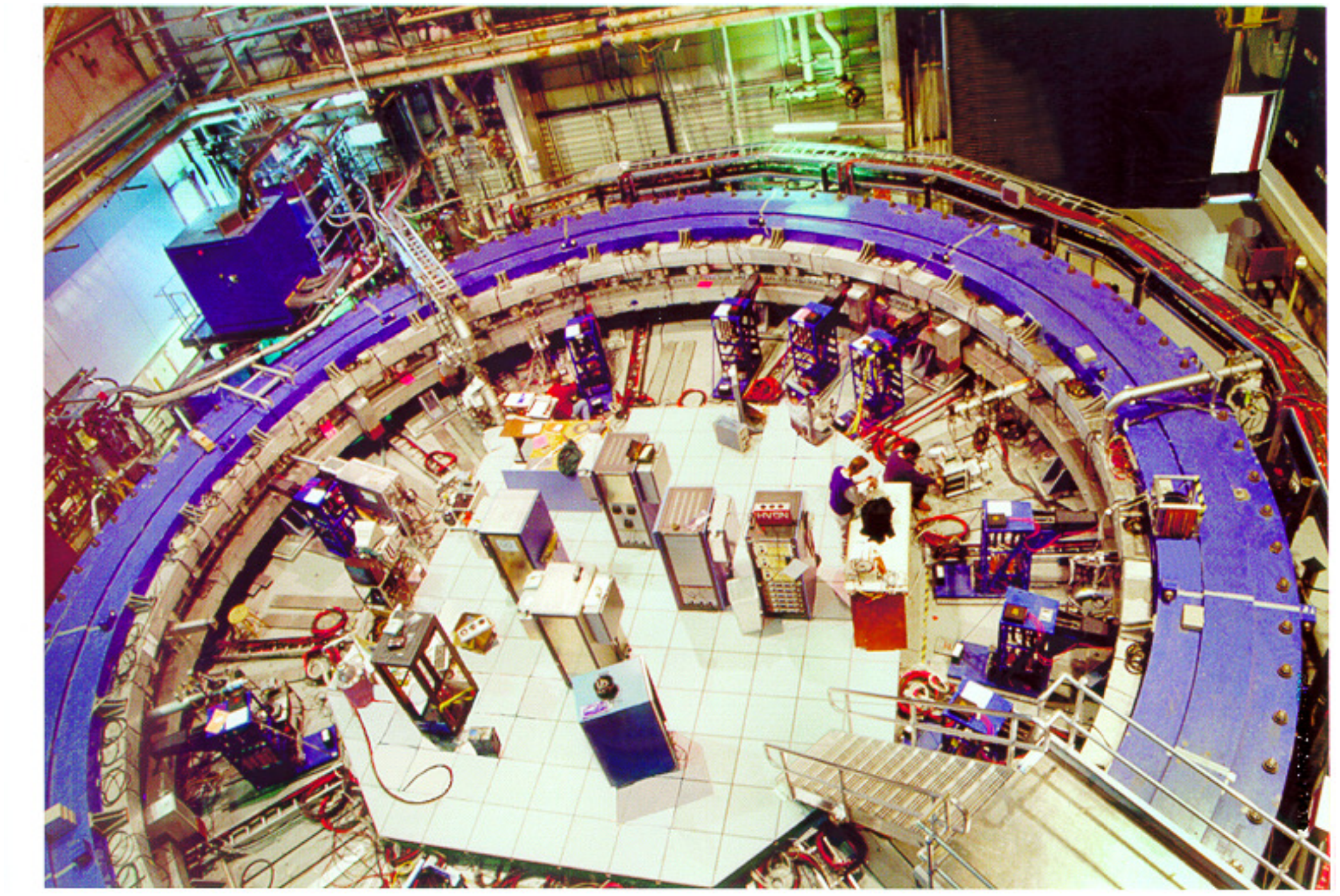}}
\subfloat[]{
 \includegraphics[width=0.5\textwidth,angle=0]{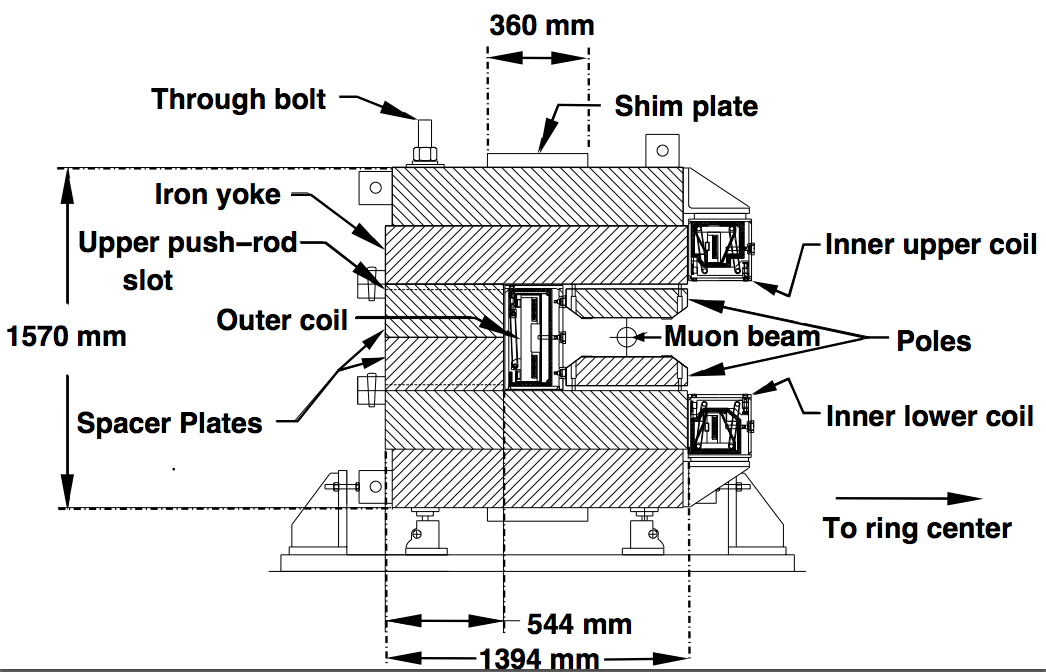}}
\caption{
  (a) A photograph of the E821 muon storage ring in the experimental hall at
  Brookhaven Laboratory. (Photograph by R. Bowman,
  courtesy of Brookhaven National Laboratory)
  (b) An elevation drawing of the storage ring cross section. (From
  Ref.~\cite{Danby:2001eh}). 
  \label{fg:BNL-ring}}
\end{center}
\end{figure}

 The storage ring yoke and pole design consisted of
 azimuthally continuous sections~\cite{Danby:2001eh}, unlike the 40 separate
 magnets in the CERN-3 design.
The as-built azimuthal gaps between yoke pieces was 
 0.8~mm, with an rms deviation of 0.2 mm.
 This small spacing between yoke pieces made it possible to eliminate the
``bumps'' in the magnetic field that were present in the 40 gaps between the
 individual magnets in the CERN-3 storage ring. An air gap between the main
 yoke and the pole pieces, decoupled the magnetic field in the storage region
 from possible non-uniformities in the yoke steel.  The precision pole pieces
 were fabricated from special steel that was continuous vacuum-cast steel
with 0.004\% carbon. The tolerance on flatness for the pole pieces 
was 25~$\mu$m, which represents 140~ppm of the magnet gap. A number of
shimming tools were built into the design of the magnet~\cite{Danby:2001eh}.
The upper and lower pole faces were machined parallel to 0.005 cm.  There
were two penetrations through the magnet yoke, for the beam to enter the
storage ring, and for the inflector magnet services. To minimize the flux
disturbance in the yoke, iron plates were added around the holes
on the outer radius side of
the yoke to compensate for the iron removed to make the penetrations.

\begin{figure}[h!]
\begin{center}
\subfloat[]{
  \includegraphics[width=0.4\textwidth,angle=0]{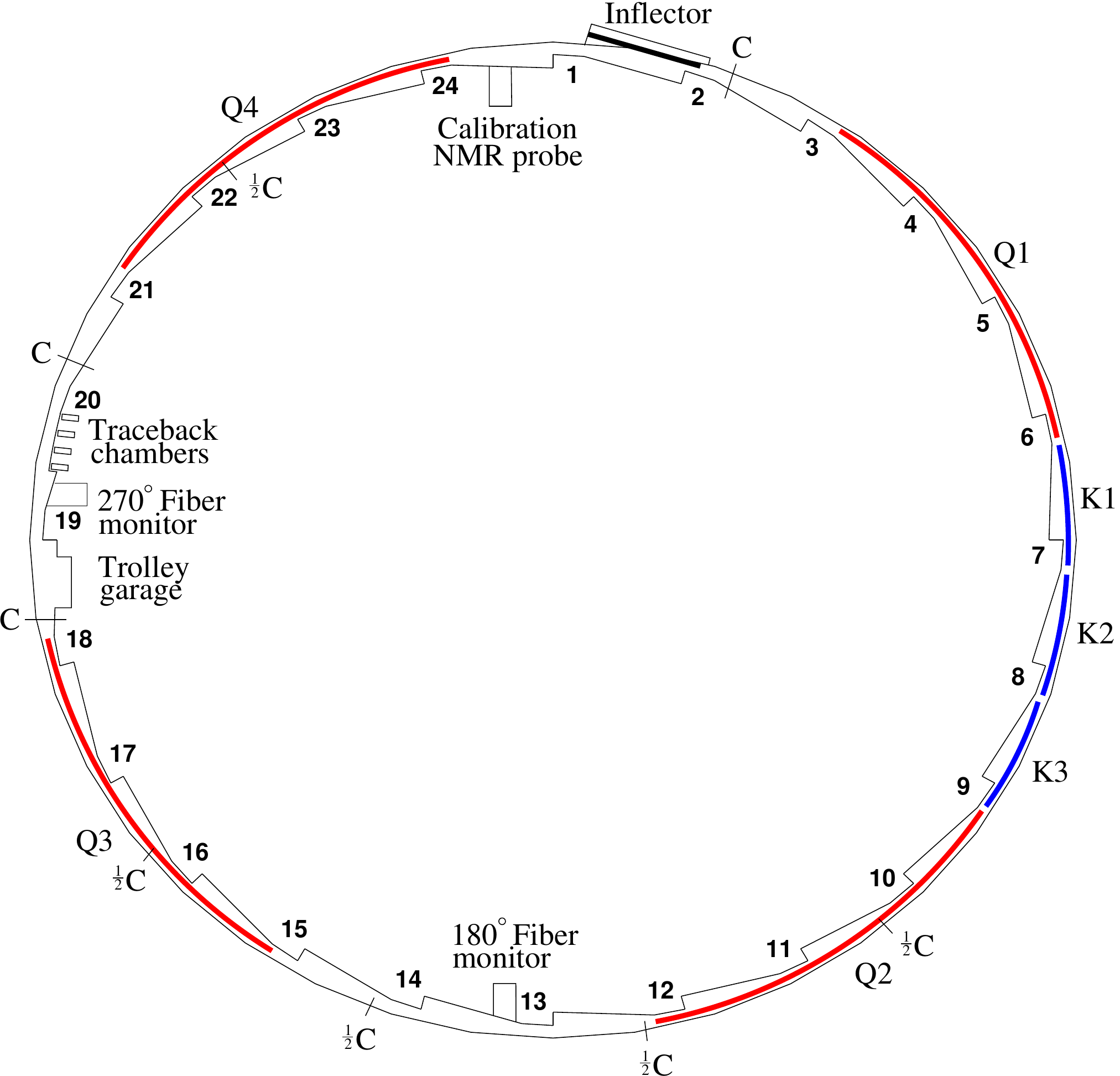}}
\subfloat[]{
 \includegraphics[width=0.5\textwidth,angle=0]{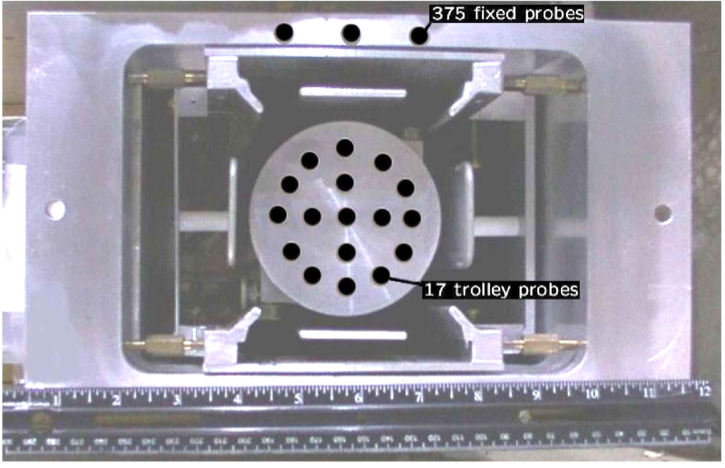}}
\caption{
  (a) The layout of the muon beam vacuum chamber showing locations of
  the electric quadrupoles (Q1 - Q4), and the fast muon kickers (K1 - K3).
  The location of the
  calorimeters are indicated by numbers at each sawtooth.
  (From Ref.\cite{Bennett:2004pv})
  (b) The inside of a vacuum chamber showing an electrostatic quadrupole, and
  the NMR field mapping trolley.  The location of the NMR probes inside the
  trolley  are indicated by black circles, as are the locations of the fixed
  NMR probes that are in the outer wall of the muon beam vacuum
  chamber. (Photo by K. Jungmann).
   \label{fg:BNL-ring-trolley}}
\end{center}
\end{figure}

The magnetic field was measured and monitored with nuclear magnetic resonance
probes~\cite{prigl:1996fpk}.  There were 378 probes (called ``fixed probes'')
placed in grooves machined into  the outside the vacuum chamber above and
beneath the muon storage region. These probes made it possible to 
track the magnetic field in the entire
storage region during the data collection
period. 
Because of magnetic gradients at the pole piece
boundaries, and only about half of these proved useful.
  An NMR trolley inside of the vacuum chamber was outfitted with 17 NMR
probes, as shown in Fig.~\ref{fg:BNL-ring-trolley}(b).  The trolley remained
in a garage during data collection.  Once every several days the beam was
turned off, and the trolley traveled around the ring measuring the field
at about 6000 points and then returned to the garage.
The trolley map was then correlated with the
fixed probe readings to determine the magnetic field seen by the muons during
data collection.

\begin{figure}[h!]
\begin{center}
\subfloat[]{
  \includegraphics[width=0.5\textwidth,angle=0]{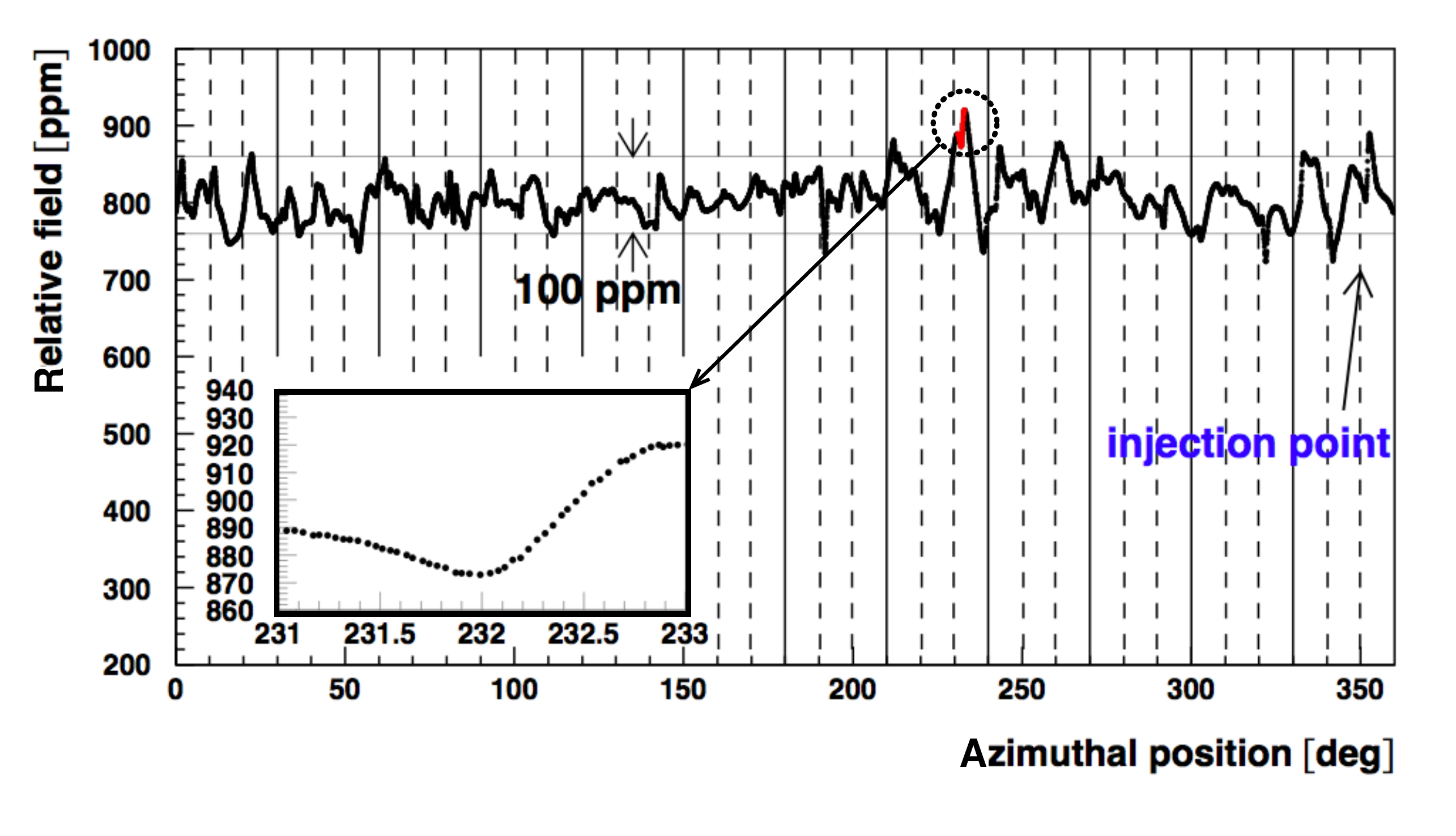}}
\subfloat[]{
 \includegraphics[width=0.3\textwidth,angle=0]{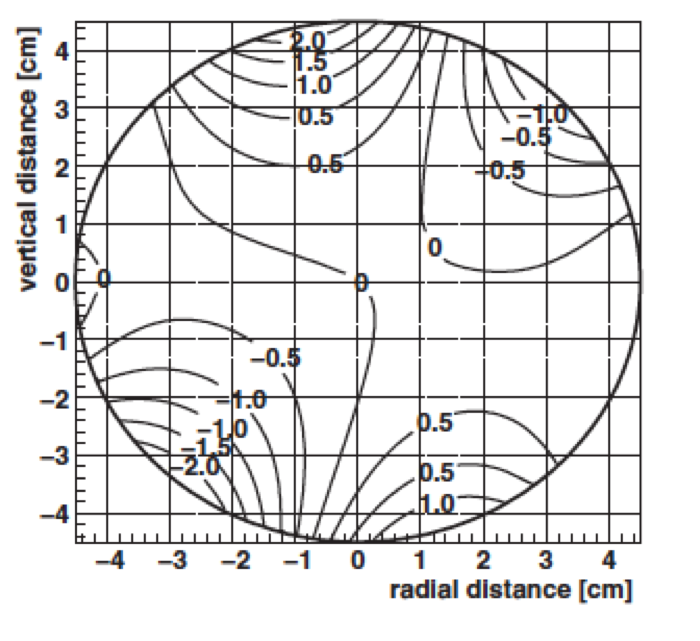}}
\caption{(a)  The magnetic dipole field from E821 as a function of
  azimuth. The inset shows the region circled on an expanded scale.
   (From Ref.\cite{Bennett:2006fi}) 
  (b) The magnetic field averaged over azimuth from E821. The contours are
  0.5~ppm. (From Ref.~\cite{Bennett:2006fi}). 
  \label{fg:B-av-field}}
\end{center}
\end{figure}

The dipole field from a sample trolley run is shown in
Fig.~\ref{fg:B-av-field}(a).
The RMS of this map is 39~ppm, with a peak-to-peak range of 230 ppm.
The field averaged over the azimuth is shown
in Fig.~\ref{fg:B-av-field}(b),  which should be compared with the CERN map in
Fig.~\ref{fg:CERN-3-field-data}.  
Since the storage ring is a weak focusing betatron,
the muons slowly sample the full
azimuthal magnetic field, so the magnetic field averaged over azimuth is the
relevant quantity.

Twenty four lead-scintillating fiber electromagnetic
calorimeters~\cite{Sedykh:2000ex} were placed 
symmetrically around the storage ring. Four acrylic lightguides carried the
light to four photomultipliers which were analog summed and then wave-form
digitized.
The muon beam vacuum chambers had a
``sawtooth'' arrangement, as shown in Fig.~\ref{fg:BNL-ring-trolley}(a). The
calorimeters, located in air, were placed into this
sawtooth step. The decay positrons exited this sawtooth through a thin
Al window before striking the calorimeter, which prevented shower losses
before the calorimeter, which was important in preserving the decay
asymmetry in the data.

\begin{figure}[h!]
\begin{center}
\subfloat[]{
  \includegraphics[width=0.5\textwidth,angle=0]{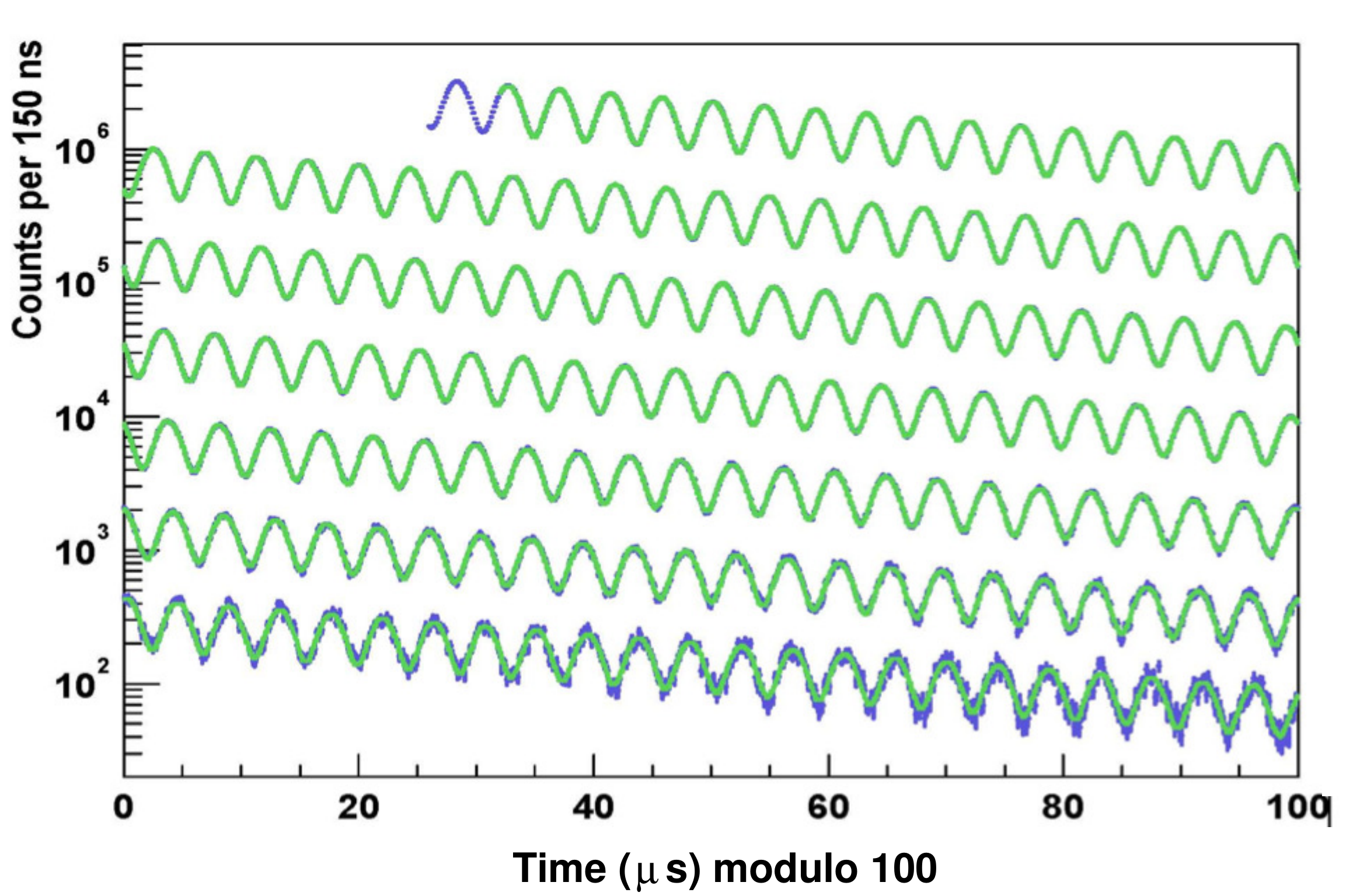}}
\subfloat[]{
 \includegraphics[width=0.35\textwidth,angle=0]{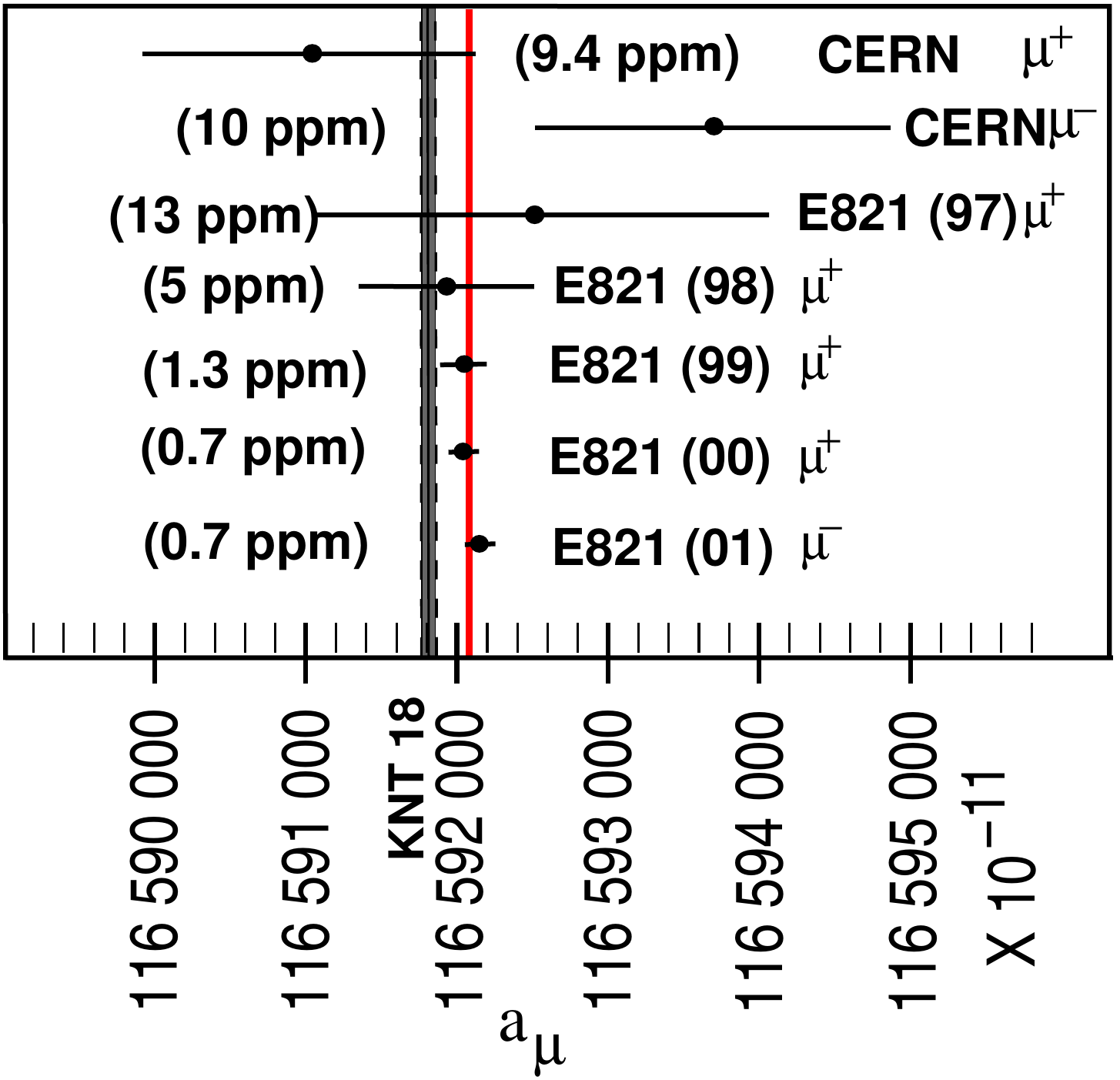}}
\caption{
  (a) The arrival time spectrum from the 2001 data
  set~\cite{Bennett:2004pv}. The blue points are data and the green curve is
  the fit to the data. The histogram contains $\simeq 3 \times 10^9$
  events. 
  (b) The measurements of $a_\mu$ from CERN-3 and from BNL E821.  The
  vertical band labeled KNT shows the Standard Model value from Keshavarzi
  et al.~\cite{Keshavarzi:2018mgv}. The thin vertical line is the combined
  average of the individual measurements. 
  \label{fg:BNL-spectrum}}
\end{center}
\end{figure}

The arrival time spectrum of electrons from the 2001 data collection period
is shown in Fig.~\ref{fg:BNL-spectrum}(a).  There are approximately
$3 \times 10^{9} $ events in this spectrum, and the precision on $\omega_a$
from this data set was 0.7~ppm.  

E821 measured both $a_{\mu^+}$ and $a_{\mu^-}$. Assuming {\it CPT} symmetry,
the final result from E821 was:
\be
  a_\mu^{\mathrm{E821}} = 116\,592\,089(54)_{stat}(33)_{syst}(63)_{tot} \times
  10^{-11}~~{(\pm 0.54\, {\rm ppm})}\, ,
  \label{eq:E821-resultI}
\ee
which is shown in Fig. \ref{fg:BNL-spectrum}(b) along with the individual
measurements. 

\section{Fermilab E989}

To clarify whether the 3.7-standard deviation difference between the Standard
Model and the E821 measurements, a new experiment was founded at
Fermilab. The experiment re-uses the storage ring magnet and the superconducting
inflector from E821, with a re-furbished NMR trolley and magnetic field
measurement system.

The new features of E989 are: 1) A pure muon beam with no hadronic component
has been designed and commissioned; 2) Segmented calorimeters consisting of
a $6 \times 9$ array of
lead fluoride crystals, which permit the observation of minimizing ionizing
particles such as muons lost from the storage ring; 3) A new fast muon kicker
using a Blumlein pulse-forming network; 4) Improved magnetic shimming that has
improved the magnetic field uniformity by a factor of two; 5) Two straw
tracker arrays inside of the vacuum chamber upstream of two of the
calorimeters; 6) More rapid rate of filling the storage ring from the
Fermilab accelerator facility.  The goal is to accumulate 21 times more data
than  E821,  to improve the systematic errors by a factor of $\simeq 3$ and
the overall uncertainty a factor of four
over E821. More details and a progress report are 
 covered in the talk by Anna Driutti at this meeting.

 \section{Summary and Conclusions}

The measurement of the muon magnetic moment spans six decades, and the story
is not over.  With the recent significant improvements in the Standard Model
value of the muon
anomaly~\cite{Davier:2017zfy,Jegerlehner:2017lbd,Keshavarzi:2018mgv}
evidence for a possible deviation between the experimental value and the
 Standard Model value
continues to grow.  The E821 result now  differs by more than three
and a half standard deviations from the Standard Model value.  Fortunately
there are two new experiments that should be able to clarify this
discrepancy. The Fermilab experiment E989, which represents the next level of
improvement in the series of ``magic $\gamma$'' storage ring experiments, is
now collecting data with the goal of a fourfold improvement over BNL E821.
 A new
experiment, E34 at J-PARC, discussed
at this conference by Tsutomu Mibe,
is developing a very different technique to measure the anomaly.

\section{Acknowledgments}

For many years
I have been deeply involved in the muon $(g-2)$ experiment E821 at
Brookhaven, and in the new Fermilab experiment E989 that is now
coming on line. I wish to thank all of my collaborators for their
contributions, and for the many things that they taught me. I especially wish
to recognize the very senior members, many of whom have died over the past
decade or so.  I
wish to acknowledge the enormous amount of knowledge that I gained from
 Francis J.M. Farley (1920 - 2018),
who played a leading role in all three CERN $(g-2)$ experiments as well as
being a collaborator on E821; Vernon
W. Hughes (1920 - 2008), founder and Co-spokesperson for E821;
Frank Krienen (1917-2008), who designed the inflector magnet and electric
quadrupoles for the CERN-3 experiment, and made many contributions to E821
including the design of the superconducting inflector magnet;
Gordon Danby (1929-2016), who came up with the brilliant magnetic design of
the E821/E989 storage ring; and Gisbert zu Putlitz, a senior collaborator
whose wisdom and leadership in E821 was essential.
I also learned many things from Emilio Picasso
(1927-2014), who was deeply involved in CERN-3, and was a friendly critic as
we prepared E821. At the very beginning of the planning for E821,
Fred Combley (1935-2001) suggested that direct muon
injection into the storage ring should be developed.  That suggestion, and its
realization by our team at Brookhaven was essential to the success of
E821. Muon injection is also a central feature of
the new Fermilab experiment.

This work was supported in part by the U.S. Department of Energy Office of
High Energy Physics.

\bibliography{BLRob-bibliography} 




\end{document}